\documentclass{iopart}

\usepackage{epsfig}
\usepackage{graphicx}

\begin{document}

\title[]{Manifold algorithmic errors in quantum computers with static internal imperfections}

\author{Murat \c{C}etinba\c{s} and Joshua Wilkie}

\address{Department of Chemistry, Simon Fraser University, Burnaby, British Columbia, V5A 1S6, Canada}
\ead{cetinbas@sfu.ca and wilkie@sfu.ca}

\begin{abstract}
The inevitable existence of static internal imperfections and residual interactions in some  quantum computer architectures result 
in internal decoherence, dissipation, and destructive unitary shifts of active algorithms. By  exact numerical simulations
we determine the relative importance and origin of these errors for a Josephson charge qubit quantum computer. In
particular we determine that the dynamics of a CNOT gate interacting with its idle neighboring qubits via native residual coupling behaves much like a perturbed kicked
top in the exponential decay regime, where fidelity decay is only weakly dependent on perturbation strength. 
This means that retroactive removal of gate errors (whether unitary or non-unitary) may not be possible, and that effective error correction schemes must 
operate concurrently with the implementation of subcomponents of the gate.
\end{abstract}


\submitto{\JPA}
\maketitle

\section{Introduction}

In this manuscript, we study the effects of one-- and two--body static flaws on a quantum controlled-NOT (CNOT) gate performed on two qubits of a larger Josephson charge-qubit quantum computer (QC) \cite{Nori,QC}. The expected external decoherence time 
for this architecture allows up to $10^{6}$ single qubit operations, and hence issues of 
internal decoherence and other intrinsic errors are of paramount concern\cite{CW2,CW3,CW4,CW5,GS,SG,BCS}. By internal decoherence we mean 
errors in the basic structure of the QC such as incorrect qubit parameters or unwanted qubit-qubit interactions, while by external decoherence
we refer to the effects of unwanted interactions with classical external fields and structures which are not part of the QC. The isolated QC core, with internal decoherence sources only, can be mapped onto a subsystem-bath scheme
wherein the active part, a two-qubit register (the subsystem), performs the CNOT gate while interacting residually with the neighboring idle qubits (the bath). We 
evolve the QC exactly from a variety of configurations: eight different initial register states, two different error generators (phase and bit-flip errors) and five different intra-bath coupling strengths and we monitor the emergence of
errors with two error quantifiers: gate purity and fidelity. These measures allow us to distinguish non-unitary errors (i.e. decoherence and dissipation) from those of unitary type (i.e. coherent shifting or distortions). 

A first brief report based on our study\cite{CW2} revealed two important effects: the suppression of non-unitary errors with increasing
intra-bath coupling, and the existence of a destructive universal coherent shift. The importance of the first effect is that one can in 
principle manipulate the idle qubit environment to improve the performance of active qubits\cite{CW2}. Thus, chaotic bath interactions may prove to be useful as an error correcting strategy, as suggested in \cite{Tess}.
The second effect revealed that flawed QCs are subject to destructive bath-induced unitary errors due to coherent shifting. While the coherent shift is a potentially harmful error source, also a focus of 
a related study\cite{CW3}, it can be used to probe internal bath dynamics and to estimate the two-body residual interaction strength.

Two unresolved mysteries connected with the time scales 
and strength of fidelity decays emerged from these initial reports\cite{CW2,CW3}. 
In \cite{CW3}, where the subsystem is a single qubit, the fidelity decayed slowly over times greater than 30 ns, but in the CNOT study discussed 
here and initially reported in \cite{CW2} the fidelity decays in about 1 ns. A natural question is why the fidelity decay should be so much worse for
the CNOT subsystem than for the single qubit subsystem, despite the fact that the magnitude of the shift was identical in the two studies? Secondly, an 
interesting and useful property of the subsystem-bath configuration in \cite{CW3} was that the fidelity was strongly dependent on the 
intra-bath residual coupling strength. In the CNOT study reported in \cite{CW2} this sensitivity almost totally vanishes for the same type of coupling operator. 

This manuscript seeks to gain some
understanding of both of these effects. Basically, we conclude that the dynamics of the CNOT gate
behaves like a perturbed kicked top\cite{Kick} in the exponential decay regime for $xx$- type error generator (bit-flip errors) and  in the Golden Rule regime for $zz$-type error generator (phase errors), while the dynamics of a single qubit subsystem of \cite{CW3} behaves like a weakly kicked
spin in the Golden Rule regime. The fact that the CNOT gate dynamics behaves like a kicked top means that the unitary errors generated cannot
be corrected after the gate is implemented, but must be corrected concurrently with the implementation of the gate. This fact may prove very useful
in future attempts to develop the necessary error correction strategies for the bath-induced unitary errors.

In our initial report \cite{CW2}, we examined the average values of error quantifiers i.e. purity and fidelity, which allowed us to compare the overall relative importance of errors generated by two different types of coupling operators. Here we also give particular attention to the individual initial states to determine whether errors show state dependencies. Small differences in errors accumulated over a 
single gate may eventually lead to much larger differences after many gates. Decoherence is a state dependent phenomena, so state selectivity is of interest especially in the case of non-unitary errors. 

Suppression of decoherence with increasing intra-bath interactions has been explained in terms of a corresponding decline in the canonical
variance of the bath coupling operator caused by the vanishing of off-diagonal matrix elements\cite{CW2}. While this explanation is correct for the
models considered in \cite{CW2,CW3}, and it is even possible to use the vanishing of off-diagonals to devise a numerical method for open quantum systems
interacting with chaotic  
environments\cite{CW4,CW5}, a decline of variance cannot satisfactorily explain the suppression of decoherence in all cases. Here we consider $zz$- as well as the native $xx$-type subsystem-bath interactions. By comparing
and contrasting the two cases we gain considerable insight into the origin of these discrepancies and we show that in the case of $zz$-type interaction,
the suppression of decoherence with increased intra-bath coupling is caused by an increasingly more Markovian character to the subsystem dynamics. 

The original motivation behind this study was to determine what types of errors emerge in the actual dynamics of a statically flawed QC.
Recent studies\cite{GS} of the statistical properties of isolated flawed QCs\cite{SG} show that sufficiently strong residual two-qubit interactions 
cause the onset of chaos and consequent dynamical thermalization of the QC core\cite{BCS}. A suggested remedy\cite{BCS} is to 
keep residual inter-qubit coupling strengths below the critical values beyond which chaos appears. However, this can be 
a problematic solution. The computation time of a gate operation is inversely 
proportional to external field strength, and longer gate operations are more likely to be decohered by external influences. 
In any case, the circumstances under which internal chaos is truly harmful are far from clear.

Perturbative studies\cite{Prosen,Frahm} wherein dynamical fluctuations in the QC Hamiltonian are modeled by random kicks show
better fidelity decay when the perturbations are chaotic. This suggests that chaos can actually stabilize the quantum motion, which results in better fidelity. 
This prediction is consistent with a number of studies\cite{Tess,Mil} of a 
central system interacting with an external self-interacting many-spin bath, but inconsistent with other
related studies\cite{Harmon}.
The controversy over whether chaos is harmful\cite{Zurek,Alicki} may soon be satisfactorily resolved. A recent
study appears to unify the various results in a picture wherein chaos is beneficial at low temperatures but harmful at 
high temperatures\cite{Sanz}.

Previous studies of flawed QCs do not address the specifics of an operating isolated QC
architecture. Nor do these models address what specifically happens to an algorithm in the presence of such flaws. An ideal gate
sequence for one architecture may be quite different from that of another architecture. Hence, one can
ask what parts of an algorithm are affected worst (e.g. one- or two-qubit gates), and whether they are irreversibly altered via internal decoherence 
or dissipation or merely coherently shifted. Such information could be essential for optimizing performance of a QC architecture, and further development of error correction schemes\cite{Brown}. Hence, there are many questions that cannot be meaningfully investigated in the 
context of random matrix or other abstract models. A closer examination of the effects of internal errors in actual QC proposals is thus warranted. 

The organization of this manuscript is as follows: In Section 2 we discuss the mathematical details of our model for an isolated 
QC including the two-qubit register on which the CNOT gate is performed. In Section 3 the parameters used in the 
calculations and the computational methods of our study are summarized. Error quantifiers are discussed in Section 4. In Section 5 we present numerically exact 
results for two error quantifiers: purity and fidelity. The observed effects are explained in Sections 6 and 7. In Section 8 we summarize our results and discuss possible extensions of our study.

\section{Isolated statically flawed QC model}

While QCs with as few as 50 qubits could be usefully employed for simulation of interesting dynamical systems such as spin-chains and
quantum maps, for more general purposes a QC will need to have thousands of qubits to be competitive with a classical computer\cite{Steane}. Universal quantum computation\cite{univ1,univ2,univ3} can be implemented by the use of external fields to induce transformations on
one-- or two--qubit subcomponents of the overall QC. Many such transformations may need to be formed concurrently.
Ideally then, an isolated QC consists of a set of qubits which are 
not self-interacting unless they are actively participating in a two-qubit gate
operation. In practice local residual interactions exist which can couple 
active qubits unintentionally to neighboring idle qubits, which can cause unavoidable internal errors on a performed quantum algorithm. A recently proposed Josephson charge-qubit QC\cite{Nori} architecture on which our study is based, for example, is 
prone to a variety of internal noise sources\cite{Nori,exp-JJ,exp-JJ2,TLS} and
residual interactions among qubits are thus very likely to exist. 

Thus, to realistically model internal errors in such a QC core
one should in principle include static two-body interactions with at least idle nearest 
neighbor qubits. Additionally, one-qubit structural defects
are also possible and will be included in our model. Of course other effects, such as dynamical fluctuations in qubit control parameters due to
noisy time-dependent electromagnetic fields, which directly affect the active qubits, are also expected. For simplicity we
neglect these dynamical effects in this initial study. 

In section 2.1 we discuss a statically flawed 
spin-bath model representing two qubits of a QC on which a CNOT
gate is performed and a set of neighboring idle qubits. The initial states of 
the QC are discussed in 2.2. Here again we avoid the complications inherent in
a discussion of imperfect initial conditions to focus solely on the effects of static flaws. Finally, we restrict our discussion
to a low temperature regime relevant to QC design\cite{Makhlin}. Our model thus remains an idealization
to some extent, but simultaneous inclusion of all possible errors would only impede 
our understanding of each error type.

\subsection{Hamiltonians}
 
The total Hamiltonian of the isolated QC is of the form
\begin{equation}
\hat{H}(t)=\hat{H}_{S}(t)+\hat{H}_{SB}+\hat{H}_{B}
\end{equation}
where $\hat{H}_{S}(t)$ is the two-qubit control Hamiltonian for the CNOT gate, $\hat{H}_{SB}$ governs interactions between the active 
and idle parts, and $\hat{H}_{B}$ is the Hamiltonian of the idle part.

We use the following control Hamiltonian to generate the required logic operations for the CNOT gate:
\begin{equation}
\hat{H}_{S}(t)=-\frac{1}{2}\sum_{i=1}^2({\cal B}_{i}^{x}(t)\hat{\sigma}_{x}^{i} +{\cal B}_{i}^{z}(t)\hat{\sigma}_{z}^{i} )
+ {\cal J}_{x}(t)\hat{\sigma}_{x}^{1}\hat{\sigma}_{x}^{2}.
\label{ctrl}
\end{equation}
\noindent
Hamiltonian (\ref{ctrl}), the basis of a Josephson charge-qubit QC\cite{Nori} proposal, allows a scalable design wherein any two charge qubits in a circuit can be effectively coupled by a 
common super-conducting inductance. In addition, the Hamiltonian (\ref{ctrl}) requires only one two-qubit operation to implement a CNOT gate. Detailed discussions on how to generate one and two-qubit gates by external manipulations of fields are given in \cite{Nori}. We discuss experimentally accessible values of control parameters for (\ref{ctrl}) in Section 3. 

Clearly the experimental manipulations \cite{Nori} required to generate Eq. (\ref{ctrl}) can induce a potential source of error. 
We need to make some simplifying assumptions regarding the implementation
of the gate in order to concentrate on errors induced by static 
internal imperfections. In what follows, we assume full control over the
dynamics of the CNOT gate. In particular, we assume that the propagation of the 
CNOT gate can be achieved with perfect square pulses\cite{TH} which could only
be approximately implemented experimentally. Moreover, we do not allow any free hamiltonian evolution by assuming that consecutive elementary gates comprising CNOT protocol can be simultaneously switched on and off. In other words, the field strengths experienced by qubits can be switched on and off instantaneously via ${\cal{B}}_{i}^{x/z}(t)={\cal{B}}_{i}^{x/z} [\Theta (t-t_{\rm on})-\Theta(t-t_{\rm off}) ]$ for $i=1,2$ and  ${\cal{J}}_{x}(t)={\cal{J}}_{x}[\Theta (t-t_{\rm on})-\Theta(t-t_{\rm off}) ]$, and these are constant in the interval $[t_{\rm on},t_{\rm off}]$. Here the superscript $x/z$ means $x$ or $z$. Thus, the full implementation of the CNOT gate can be achieved in nine steps. These steps consist of Schr\"{o}dinger evolutions in time intervals $[\tau_{i},\tau_{i+1}]$ for $i=0,..,8$. The switching times for the nine square pulses of the CNOT gate, and the corresponding 
active Hamiltonian in each time interval, are summarized in Table I. The unitary operator 
governing the gate can thus be written as
\begin{eqnarray}
\hat{U}_{  {\rm CNOT} } &=&
\hat{U}(\tau_{9},\tau_{8})\hat{U}(\tau_{8},\tau_{7})
\hat{U}(\tau_{7},\tau_{6})\hat{U}(\tau_{6},\tau_{5})
\hat{U}(\tau_{5},\tau_{4}) \nonumber \\
&\times& \hat{U}(\tau_{4},\tau_{3}) 
\hat{U}(\tau_{3},\tau_{2})\hat{U}(\tau_{2},\tau_{1})\hat{U}(\tau_{1},\tau_{0}).
\end{eqnarray}
At this stage in the development of our model a dynamical simulation
would show perfect gate fidelity. We now include interaction with the 
neighboring idle qubits.

\begin{table}
\caption{\label{sequence} Switching times and active Hamiltonians used to \\
implement the CNOT gate.}
\begin{center}
\begin{tabular}{ccc}
\hline  \tabularnewline
Switching Intervals~~~~~~~~~~~~~~~~~~~~~~&~~~Active Hamiltonian 
\tabularnewline \tabularnewline
\hline 
\tabularnewline
$[\tau_{0}=0, \tau_{1} = \pi / (2{\cal{B}}^{z})]~~~~~~~~~~~~~~~~~~~~~~$
&
$-\frac{1}{2}{\cal{B}}^{z}\hat{\sigma}_{z}^{2}$ \vspace{0.05in}
\tabularnewline
$[\tau_{1}, \tau_{2} = \tau_{1}+ \pi / (2{\cal{B}}^{x})]~~~~~~~~~~~~~~~~~~~~~$
&
$-\frac{1}{2}{\cal{B}}^{x} \hat{\sigma}_{x}^{2}$  \vspace{0.05in}
\tabularnewline
$[\tau_{2}, \tau_{3} = \tau_{2}+ \pi / (2{\cal{B}}^{z})]~~~~~~~~~~~~~~~~~~~~~$
&
$+\frac{1}{2}{\cal{B}}^{z}\hat{\sigma}_{z}^{2}$  \vspace{0.05in}
\tabularnewline
$[\tau_{3}, \tau_{4} = \tau_{3}+ \sqrt{2} \pi / (2{\cal{B}}^{z})]~~~~~~~~~~~~~~~~~$
&
$-\frac{1}{2}{\cal{B}}^{z}\sum_{i=1}^{2}(\hat{\sigma}_{z}^{i}+\hat{\sigma}_{x}^{i})$ \vspace{0.05in}
\tabularnewline
$[\tau_{4}, \tau_{5} = \tau_{4}+ \pi / (4{\cal{J}}_x)]~~~~~~~~~~~~~~~~~~~~~$
&
${\cal J}_{x}(-\hat{\sigma}_{x}^{1}-\hat{\sigma}_{x}^{2}+\hat{\sigma}_{x}^{1}\hat{\sigma}_{x}^{2})$ \vspace{0.05in}
\tabularnewline
$[\tau_{5}, \tau_{6} = \tau_{5}+ \sqrt{2} \pi / (2{\cal{B}}^{z})]~~~~~~~~~~~~~~~~~$
&
$+\frac{1}{2}{\cal{B}}^{z}\sum_{i=1}^{2}(\hat{\sigma}_{z}^{i}+\hat{\sigma}_{x}^{i})$ \vspace{0.05in}
\tabularnewline
$[\tau_{6}, \tau_{7} =\tau_{6}+ \pi / (2{\cal{B}}^{z})]~~~~~~~~~~~~~~~~~~~~~$
&
$-\frac{1}{2}{\cal{B}}^{z}\hat{\sigma}_{z}^{2}$ \vspace{0.05in}
\tabularnewline 
$[\tau_{7}, \tau_{8} =\tau_{7}+ \pi / (2{\cal{B}}^{x})]~~~~~~~~~~~~~~~~~~~~~$
&
$+\frac{1}{2}{\cal{B}}^{x}\hat{\sigma}_{x}^{2}$ \vspace{0.05in}
\tabularnewline
$[\tau_{8}, \tau_{9} =\tau_{8}+ \pi / (2{\cal{B}}^{z})]~~~~~~~~~~~~~~~~~~~~~$
&
$+\frac{1}{2}{\cal{B}}^{z}\hat{\sigma}_{z}^{2}$
\tabularnewline
\tabularnewline
\hline
\end{tabular}
\end{center} 
\end{table}

We investigate two different types of errors separately\cite{TH}. Bit--flip 
errors are modeled by an interaction Hamiltonian of the form
\begin{equation} 
\hat{H}_{SB}= ( \hat{\sigma}_{x}^{1}+\hat{\sigma}_{x}^{2} )\sum_{i=3}^{N+2}\lambda_{i}^{x}\hat{\sigma}_{x}^{i},
\label{bit-flip}
\end{equation}
while phase errors are modeled by 
\begin{equation}
\hat{H}_{SB}=( \hat{\sigma}_{z}^{1}+\hat{\sigma}_{z}^{2} )\sum_{i=3}^{N+2}\lambda_{i}^{z}\hat{\sigma}_{z}^{i}.
\label{phase}
\end{equation}
Although solid and condensed phase QC proposals inherit a variety of physical interactions to couple qubits, and accordingly two-qubit residual interactions of $xx$-, $zz$-, $yy$-, or $xy$-type are possible error generators, we expect only $xx$-type residual interactions for the Josephson charge-qubit QC\cite{Nori} under investigation. This is because, for $xx$-type interactions, 
the active and idle parts will be fully symmetric. However, consideration of this second asymmetric $zz$-type of coupling will greatly improve our understanding
of the native symmetric dynamics.

The collective dynamics of the idle qubits is modeled via an $N$--body 
self-interacting qubit--bath Hamiltonian,
\begin{eqnarray}
\hat{H}_{B} = -\frac{1}{2} \sum_{i=3}^{N+2} \left( B_{i}^{x}\hat{\sigma}_{x}^{i} 
+ B_{i}^{z} \hat{\sigma}_{z}^{i} \right)
+ \sum_{i=3}^{N+1}\sum_{j=i+1}^{N+2} J_{x}^{i,j} \hat{\sigma}_{x}^{i}\hat{\sigma}_{x}^{j}.
\label{Hb}
\end{eqnarray}
\noindent
Hamiltonians similar to (\ref{Hb}) were the focus of a number of previous studies\cite{GS,SG,BCS} but obviously it is not the most general Hamiltonian for a qubit based QC.

Static one-body fluctuations are modeled by randomly and uniformly sampling coefficients 
from the interval
\begin{equation}
B_{i}^{x/z} \in \ [B_{0}^{x/z} \! -\delta/2, \: B_{0}^{x/z} \! +\delta/2].
\end{equation} 
We assume that the idle qubits are similar to the active qubits since they all are the components of the same QC. Thus, the average values $B_{0}^{x/z}$ of the distribution represent the native qubit dynamics in the absence of flaws. In other words, the idle qubits 
differ from the active qubits by a static noise characterized by a detuning parameter $\delta$. 

In modeling two-body residual interactions i.e. system-bath interactions and intra-bath interactions, we have followed the Refs. \cite{GS,SG}. So, we have sampled the coupling coefficients randomly and uniformly from $J_{x}^{i,j} \in [-J_x,\ J_x]$, and $\lambda_{i}^{x/z} \in [-\lambda,\ \lambda]$. In fact we used the same set of $\lambda_i$ values for both $xx$ and $zz$ coupling. In previous studies \cite{GS,SG} the QC core is assumed to be a two-dimensional lattice of qubits interacting via nearest-neighbor qubit-qubit interactions. While we use the same type of distribution for qubit-qubit interactions we also allowed all possible qubit-qubit interaction in our bath Hamiltonian [see Eq. (\ref{Hb})]. This is because the charge-qubit QC proposal \cite{Nori} under investigation permits all pairwise qubit couplings in principle and so the residual interactions among all qubit pairs are likely to exist. Therefore, we expect that the qubit-qubit interactions in our model mimic short as well as long range interactions. Nevertheless, the type of the distribution and allowed interactions in our model should still be considered as an idealization. This is because, for example, we did not explicitly take into account spacial dependencies of the noise due to the location of qubits in the circuit. In an actual experiment, the magnitude and type of the noise should be at least partially intrinsic to the particular experimental conditions and physical set-up. Nevertheless, we try to compensate for this idealization of our model by employing a large range of parameters in our numerical simulations.  
 
\subsection{Initial conditions}

We implement the CNOT gate for eight different initial pure states 
\begin{equation}
\hat{\rho}_{S}^{i}(0) =|\psi_{0}^{i}\rangle\langle \psi_{0}^{i}|
\end{equation}
of the active qubits, where $|\psi_{0}^{i}\rangle$ are chosen from two distinct groups: $\{|00\rangle,|01\rangle,|10\rangle,|11\rangle\}$ is a set of four standard basis states, and
$\{(|00\rangle + |11\rangle)/\sqrt{2}, (|00\rangle - |11\rangle)/\sqrt{2}, (|01\rangle + |10\rangle)/\sqrt{2}, (|01\rangle - |10\rangle)/\sqrt{2}\}$ is a set of four Bell states. 

We assume that the bath has been idle long enough to have thermalized
so that the initial states of the computer are of the form
\begin{equation}
\hat{\rho}^i (0) = \hat{\rho}_{S}^i (0) \otimes \hat{\rho}_{B} (0)\label{initial}
\end{equation}
where $\hat{\rho}_{B}(0)$ is the canonical bath density at equilibrium given by
\begin{equation} 
\hat{\rho}_{B} (0) = (1/Q) \exp{(-\hat{H}_{B}/kT)}
\end{equation}
where $Q$ is the partition function
\begin{equation} 
Q = {\rm Tr}_{B} [ \exp{(-\hat{H}_{B}/kT)} ].
\end{equation}
Direct product states like (\ref{initial}) are not easily achieved in practice, but inclusion of 
the effects of imperfect initial 
conditions would greatly complicate our study. Moreover, we wish to observe the dynamical emergence of
errors from residual static internal interactions, and the presence of imperfect initial conditions would 
only cloud the matter.

\section{Numerics}

Our numerical simulations are based on a recent charge qubit QC proposal\cite{Nori} 
for which the experimentally realizable control parameters are ${\cal{B}}^{x}\!={\cal{B}}^{z}\!=200$ mK. Hence, a typical switching time for the one-qubit gate operations is of order $\hbar/2{\cal{B}}^{z} \sim 0.1$ ns. Two-qubit switching times are however 10 times longer. The total gate time for the CNOT gate is then about $\tau_{9}=1.129$ ns. The relevant temperature is $50$ mK\cite{Makhlin}. While achieving this low temperature, necessary for coherent quantum control, might be an experimental burden it leads to significant computational advantages for exact propagation.

\subsection{Parameters}

For computational convenience we scale the parameters of the control Hamiltonian in units of $\epsilon=200$ mK. The one- and two-qubit control parameters are thus ${\cal{B}}^{x}\!={\cal{B}}^{z}\!=1$, and 
${\cal{J}}_{x}=0.05$, respectively, and $kT=0.25$. The other parameters that define the idle qubits are $B_{0}^{x}=B_{0}^{z}=1$ and $\delta=0.4$. We considered a number of $J_{x}$ values in
order to explore the emergence of chaos: $J_{x}=0.05, \: 0.25, \: 0.50, \:1.00, \:2.00$. The subsystem-bath interaction strength was set equal
to the two-qubit control parameter, and thus $\lambda=0.05$ for both bit-flip and phase errors.

\subsection{Simulations}

The fact that we are in the low temperature limit allows us to write the initial bath state as 
\begin{equation}
\hat{\rho}_{B}(0) =\sum_{n=1}^{n_{cut}} |\phi_{n}^{B} \rangle \frac{e^{-E_{n}/kT}}{Q^{\prime}} \langle \phi_{n}^{B} |\label{sum}
\end{equation}
where the sum is over the thermally populated, lowest energy, eigenstates of the bath. Note that
\begin{equation}
\hat{H}_{B} | \phi_{n}^{B} \rangle = E_{n} | \phi_{n}^{B} \rangle
\end{equation}
and
\begin{equation}
Q^{\prime}=\sum_{n=1}^{n_{cut}}\exp{(-E_{n}/kT)}
\end{equation} 
where the energies $E_n$ are ordered such that $E_n\leq E_m$ if $n<m$.
Thus $n_{cut}$ is not necessarily the total number of
eigenvalues but rather a cutoff chosen so that states $n_{cut}+1$ and higher are unoccupied. A Lanczos algorithm\cite{ARPACK} was employed for the exact diagonalization of bath Hamiltonian (\ref{Hb}) for $N=10$ idle bath qubits and in this low-- temperature regime $n_{cut}=20$ was sufficient. Hence, our QC core can
be viewed as a qubit pair surrounded by idle nearest neighbors in a two dimensional circuit. In an actual experimental setup $\lambda_i$ may have a systematic spatially dependent part in addition to the noise component considered here. We neglect this for simplicity.

The time evolved density matrix of the computer is then exactly expressed as 
\begin{equation}
\hat{\rho}^i (t) = \sum_{n=1}^{n_{cut}} |\Psi_{n}^i(t)\rangle \frac{e^{-E_{n}/kT}}{Q^{\prime}} \langle \Psi_{n}^i(t) |
\end{equation}
where $|\Psi_{n}^i(t)\rangle$ obeys the Schr\"{o}dinger equation 
\begin{equation}
d|\Psi_n^i(t)\rangle/dt=-(i/\hbar)\hat{H}(t)|\Psi_n^i(t)\rangle\label{int}
\end{equation}
with initial conditions $|\Psi_n^i(0)\rangle=|\psi_{0}^{i}\rangle\otimes |\phi_{n}^{B} \rangle$. Here $i=1,\dots, 4$ labels the initial
state of the gate for both standard basis states and Bell states. The numerical integrations of (\ref{int}) were performed by an explicit variable-step-size Runge--Kutta method\cite{RK} of order $8$. 

\section{Error quantifiers}

The quantity of primary interest is the reduced density of the active degrees of freedom, 
$\hat{\rho}_S(t)$, obtained by tracing out the degrees of 
freedom of the idle bath qubits, i.e. 
\begin{eqnarray}
\hat{\rho}_S(t)={\rm Tr}_{B}[\hat{\rho}(t)].\nonumber
\end{eqnarray}
The reduced density supplies all necessary probabilistic information about the open dynamics of the CNOT gate. Hence, once the reduced density is known, the quality of gate implementation can readily be assessed by standard error quantifiers. We employ two error quantifiers for our assessment: purity and fidelity. Non-unitary internal errors due to decoherence and dissipation are quantified by using purity since the purity is insensitive to unitary effects. Fidelity reflects all sources of error. Hence, a large deviation between the purity and fidelity can be used as an indicator of unitary errors due to the coherent shifting process.  

Purity, also known as linear entropy, is defined by the trace of the square of the reduced density operator,
\begin{equation}
\label{pur}
{{\mathcal P}}(t)={\rm Tr}_S [\hat{\rho}_{S}^{2}(t) ],
\end{equation}
and it gives a measure of how close the reduced density stays to a pure state. Most if not all technologies which rely on 
quantum interference employ pure states exclusively for their implementation. Deviations from perfect purity are however
inevitable in practice and hence some measure of the extent of this deviation is needed. Pure states have a purity of one
and mixed states have purities less than one.

Gate fidelity can be calculated from the reduced density using
\begin{equation}
\label{fid}
{\mathcal F}(t) = {\rm Tr}_S[\hat{\rho}_{S}(t)\hat{\rho}_{S}^{ideal}(t)]
\end{equation}
where $\hat{\rho}_{S}^{ideal}(t)$ is the dynamics obtained in the absence of residual interactions with the idle qubits, i.e.,
\begin{equation} 
\hat{\rho}_{S}^{ideal}(t)=\hat{U}_{\rm CNOT}(t)\hat{\rho}_{S}(0)\hat{U}_{\rm CNOT}^{\dagger}(t).
\end{equation}
The desired value of the fidelity would be unity
at all times in the absence of coupling to the idle bath qubits.

\section{Results}

In our study we will not attempt to distinguish between decoherence and dissipation since the energy of the two-qubit register is always
changing due to the external manipulations of the control Hamiltonian. However, the distinction between non-unitary ``decoherence and dissipation''
and unitary ``coherent shifting'' is important since unitary errors are far more easily corrected.

In Figures \ref{pur-xx-f}-\ref{pur-zz-b} we examine non-unitary errors as measured via purity which we discuss below in section 5.1. Figures
\ref{pur-xx-f}-\ref{pur-zz-f} focus on standard initial states. In Figure \ref{pur-xx-f} we examine dynamics with $xx$-type coupling, while in
Figure \ref{pur-zz-f} the coupling is of $zz$-type. Each figure shows results for four initial states; the subfigure (a) is reserved for $|00\rangle$, the subfigure (b) for $|01\rangle$, the subfigure (c) for $|10\rangle$, and  the subfigure (d) for $|11\rangle$. Analogous quantities are 
reported in Figures \ref{pur-xx-b}-\ref{pur-zz-b} for Bell type initial states: the subfigure (a) is reserved for $(|00\rangle+|11\rangle)/\sqrt{2}$,  the subfigure (b) for $(|00\rangle-|11\rangle)/\sqrt{2}$, the subfigure (c) for $(|01\rangle+|10\rangle)/\sqrt{2}$, and the subfigure (d) for $(|01\rangle-|10\rangle)/\sqrt{2}$.

Figures \ref{fid-xx-f}-\ref{fid-zz-b} examine unitary errors as measured via fidelity which we discuss below in section 5.2. The subfigure
states are organized in the same way as for purity. Figures \ref{fid-xx-f} and \ref{fid-xx-b} are for $xx$-type coupling while 
\ref{fid-zz-f} and \ref{fid-zz-b} are for $zz$ coupling.

For each initial condition the purity and fidelity are shown for five different values of the intra-bath coupling $J_{x}$. We have also labeled the times at which the various manipulations involved in the gate are initiated.

We have performed ten different realizations of the QC. The results presented in this section are only 
for a single realization but it is a typical flawed QC. However, we have encountered some exceptional 
realizations. In some cases the bath density of states may increase with increasing $J_{x}$. This can cause accidental near degeneracies in the low energy spectrum of the bath. The number of thermally and 
dynamically populated bath states can then increase with increasing $J_{x}$. This then results in an
anomalous increase of decoherence with $J_{x}$. In some QC architectures (e.g. symmetric $xy$-models)
not considered here this is the dominant effect.

\subsection{Non-unitary errors from decoherence and dissipation}

\begin{figure}
\centering
\includegraphics[scale=0.35]{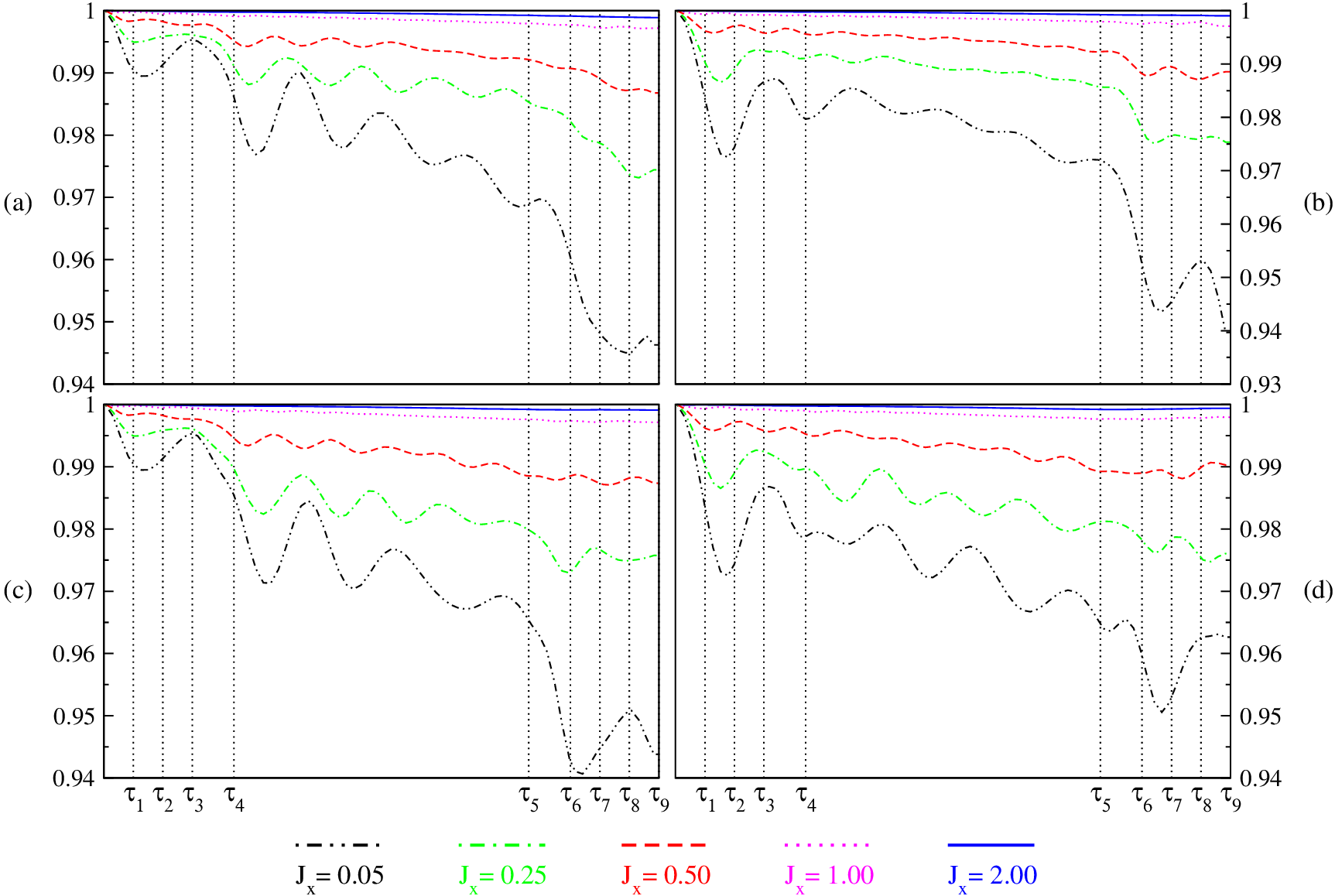}
\caption{(Color online) Purity $\mathcal{P}(t)$ vs. time for $xx$ bit--flip errors calculated for initial standard basis states. $\tau_9=1.13$ ns.} 
\label{pur-xx-f}
\end{figure}

\begin{figure}
\centering
\includegraphics[scale=0.35]{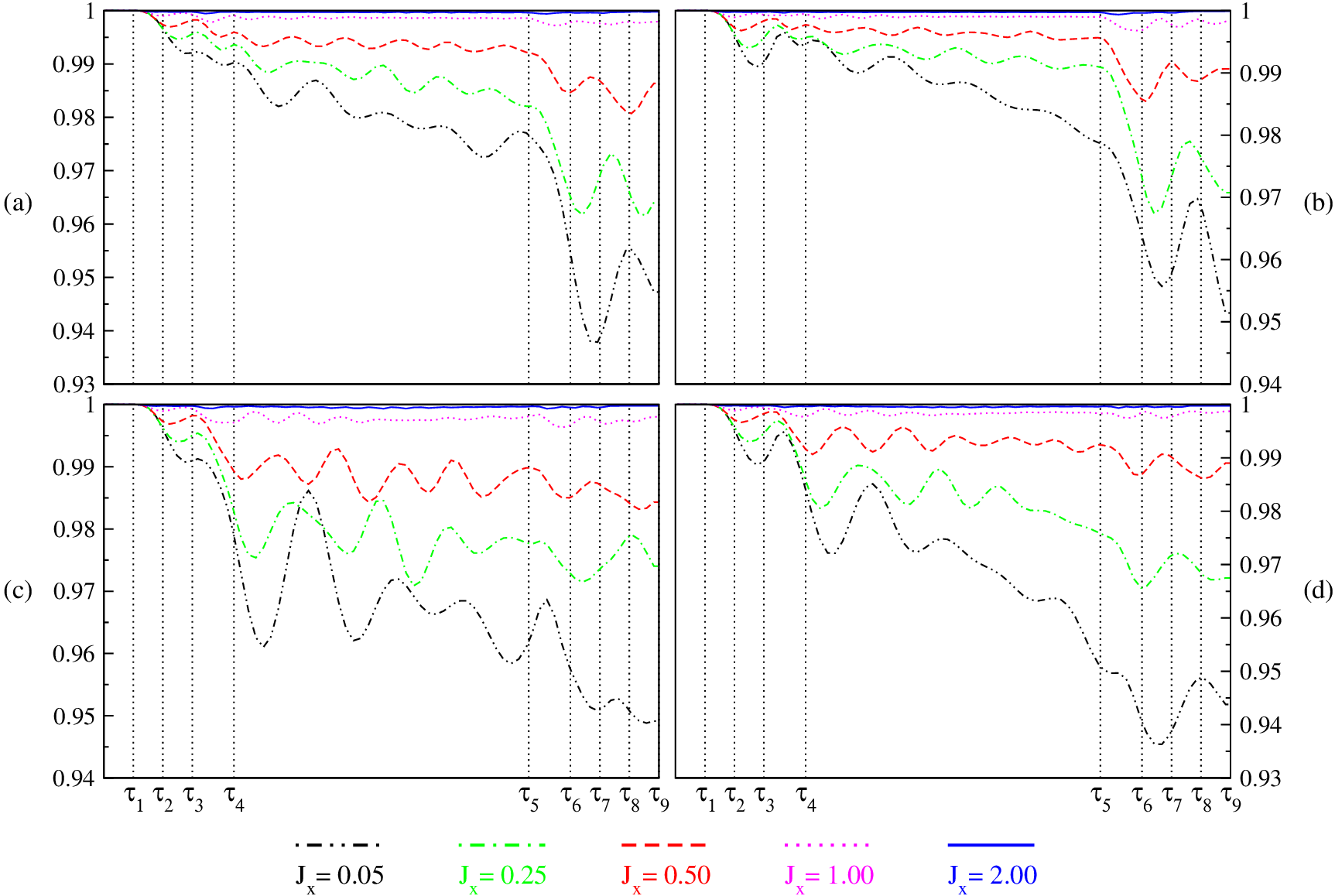}
\caption{(Color online) Purity $\mathcal{P}(t)$ vs. time for $zz$ phase errors calculated for initial standard basis states. $\tau_9=1.13$ ns.} 
\label{pur-zz-f}
\end{figure}

\begin{figure}
\centering
\includegraphics[scale=0.35]{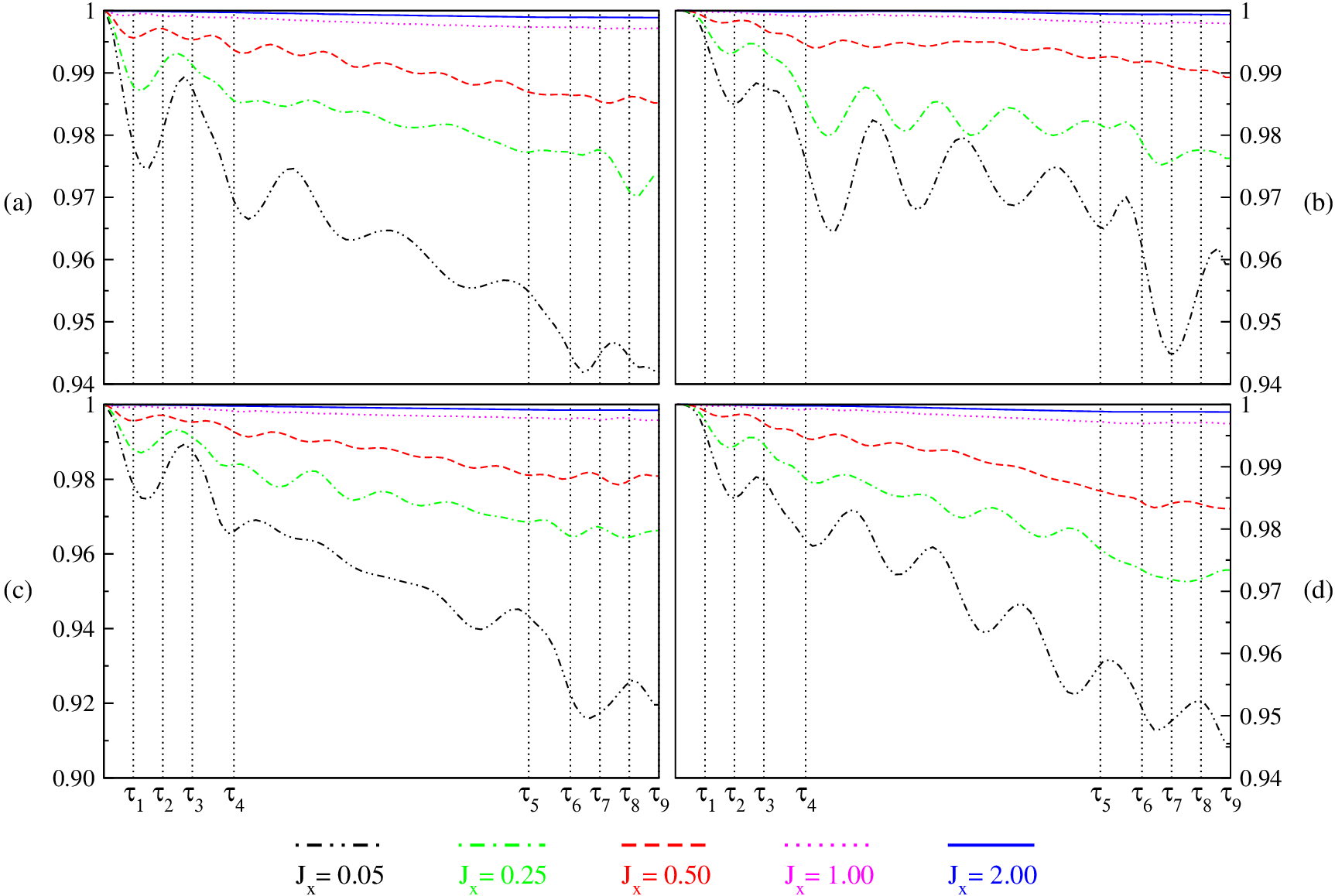}
\caption{(Color online) Purity $\mathcal{P}(t)$ vs. time for $xx$ bit--flip errors calculated for initial Bell states. $\tau_9=1.13$ ns.} 
\label{pur-xx-b}
\end{figure}

\begin{figure}
\centering
\includegraphics[scale=0.35]{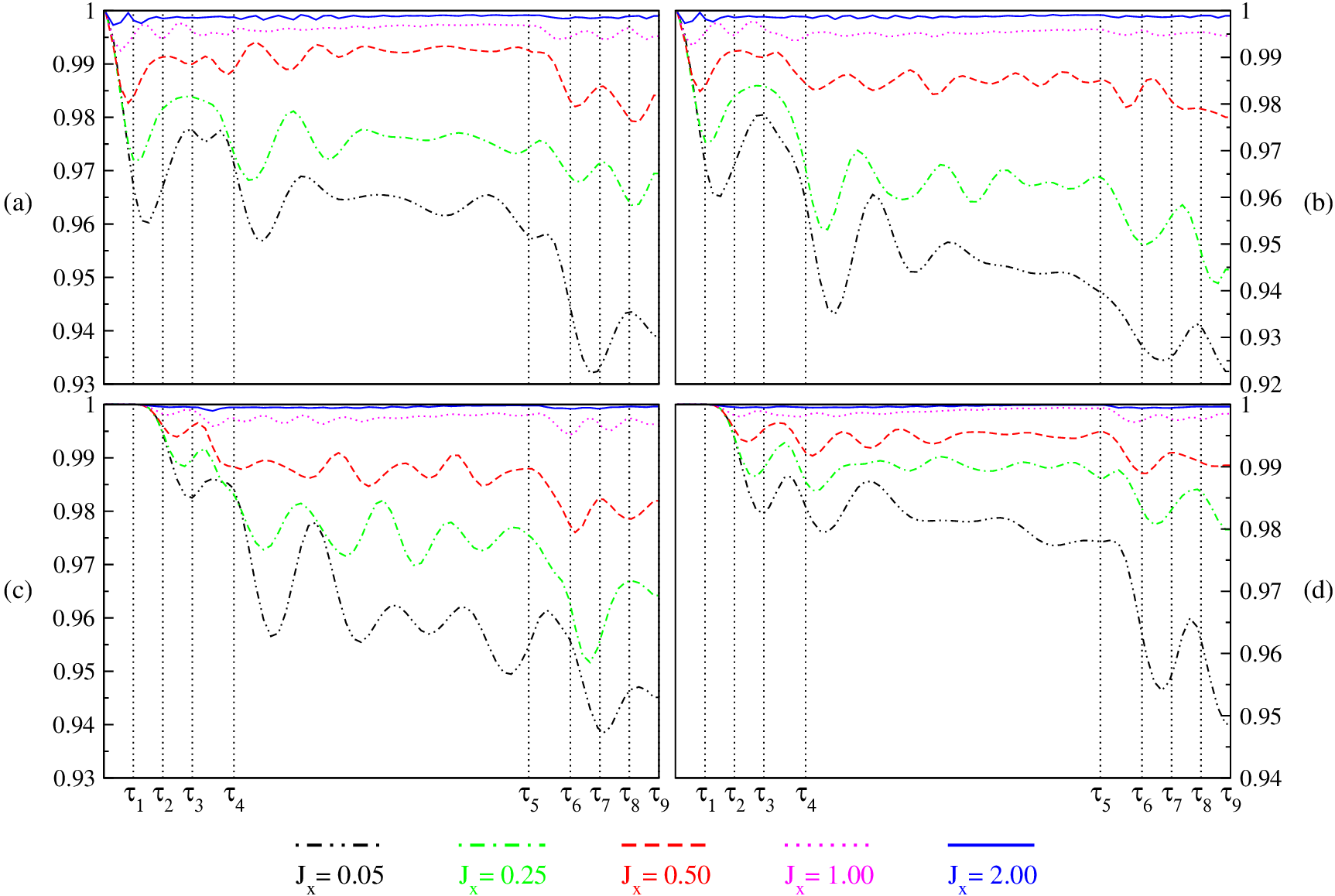}
\caption{(Color online) Purity $\mathcal{P}(t)$ vs. time for $zz$ phase errors calculated for initial Bell states. $\tau_9=1.13$ ns.} 
\label{pur-zz-b}
\end{figure}

We plot the purity vs time for $xx$-type coupling in Figure \ref{pur-xx-f} for standard initial basis states.
In Figure 1(a) we see five different curves representing the different values of $J_x$. The lower the $J_x$ the greater
is the departure from the ideal value of 1. For a near integrable bath (i.e. $J_x=0.05$) we see the greatest impurity and so
the highest non-unitary error. As $J_x$ is increased this error is systematically reduced until by the time $J_x=2$ there is
almost no non-unitary error. For the lower values of $J_x$, oscillations are observed which are indicative of the 
presence of memory effects in the dynamics. These effects suggest that integrable baths are more non-Markovian and cause
more decoherence and dissipation than chaotic baths. The other subfigures for the other standard initial states are
qualitatively similar. Clearly, however, there are some quantitative differences which are indicative of a degree of 
state specificity.

The early dynamics $t\le \tau_4$ are similar in (a) and (c), and in (b) and (d) for $J_x=0.05,0.25,0.50$. But (a) and (b) are quite different
on this time scale. We see bunching of the curves at $\tau_3$ in (a) and (c), but clearly separated
curves in (b) and (d) at the same time. While the short time similarities of (a) and (c) continue throughout the
dynamics, (b) and (d) then evolve rather differently. We see pronounced oscillations in the near integrable curves
of (d) but those of (b) are more monotonic. For $J_x=0.05$ the purity at $\tau_9$ is roughly .935 in (b) and .9625 in (d)
which is quite a big difference. Thus, there is clearly a degree of non-unitary state specificity for $xx$-type coupling
for the smaller $J_x$ values. The $J_x=1.00,2.00$ curves show almost no state specificity.

In Figure 2 we plot dynamics for the same initial states but for $zz$-type coupling. Here the plots again show an 
improvement in purity as $J_x$ increases. The magnitude of the errors is also quite similar to that for $xx$-type coupling. 
This rough similarity is quite surprising, as we will show later that the origins of the effect are different for the 
two cases. The early dynamics $t\le \tau_4$ are similar in (a) and (c), and in (b) and (d)
for the lowest $J_x$ values. There are no strong similarities in the dynamics of any subfigures after $\tau_4$.
Here state specificity appears quite strong with most divergence taking place during the long two qubit gate. The 
highest two $J_x$ values show little state specificity.

In Figure 3 we return to $xx$-type coupling but for Bell type initial states. Again we see suppression of errors with 
increasing $J_x$. For $J_x=0.05$ we see worse decoherence than we have yet encountered, but otherwise the dynamics
is qualitatively similar to that for the standard states. State specific effects are slightly less pronounced than in Figure 1.

Figure 4 shows that the switch from standard states to Bell states for phase errors is not dramatic except for (a) and (b) where the short time
dynamics is quite different. 

In all figures non-negligible deviations of the purity from the theoretically desirable limit of 0.99999 \cite{Loss} are observed for the experimentally relevant two-qubit coupling strength $J_{x}=0.05$. Recall that the number of idle bath qubits directly participating in the decoherence process is relatively low, i.e. $N=10$. This number could be higher for larger QCs, in three dimensional circuits for example. Hence, internal decoherence can be a matter of concern in a flawed QC core. On the other hand, it is also very clear that increasing the residual interaction strength $J_{x}$ causes rapid reduction of decoherence. Hence the bath chaos stabilizes the gate implementation by causing an increase in the purity. For the strongest coupling case $J_{x}=2$ the effect of decoherence almost completely vanishes for both types of coupling operators and all initial conditions. This suggests that induced bath chaos may serve as an error correcting strategy. However, while such strong bath couplings are certainly desirable they may not be experimentally accessible for this particular architecture \cite{Nori} within today's technological limits \cite{Nori} [{\it see} Section 3 for the experimentally accessible control parameters]. Nevertheless, the general effect we observe is systematic, and should be observable in the experimentally accessible regime.
  
Overall the decrease of the purities with time appears qualitatively similar for all initial conditions, but Bell states perform slightly worse than standard basis states. This should not be surprising. Since Bell states are special correlated states, they are more fragile to the destructive effects of decoherence. Performance with respect to the type of coupling operator is also quite similar in all cases. Overall, the purity decays are of comparable magnitude for both bit-flip and phase type couplings. However, intrinsic decoherence due to particular initial conditions and coupling operators is also seen. For example, decoherence by bit-flip type coupling affects the system during the first gate operation for all initial conditions. In the case of phase type coupling, however, the errors do not emerge until the second gate operation for all standard basis states (as seen in Figure \ref{pur-zz-f}) and for two of the Bell states (seen in Figures \ref{pur-zz-b}(c) and  \ref{pur-zz-b}(d)). The decoherence free dynamics observed in these cases is due to the fact that the first gate operation commutes with the coupling operator and the initial states are eigenvalues of both.  

Decoherence is clearly a state dependent phenomenon. The slight state specificity we discuss above cannot be dismissed as negligible because the
small effects observed in this one gate could get amplified over time during other gates. 

\subsection{Unitary errors from coherent shifting}

\begin{figure}
\centering
\includegraphics[scale=0.35]{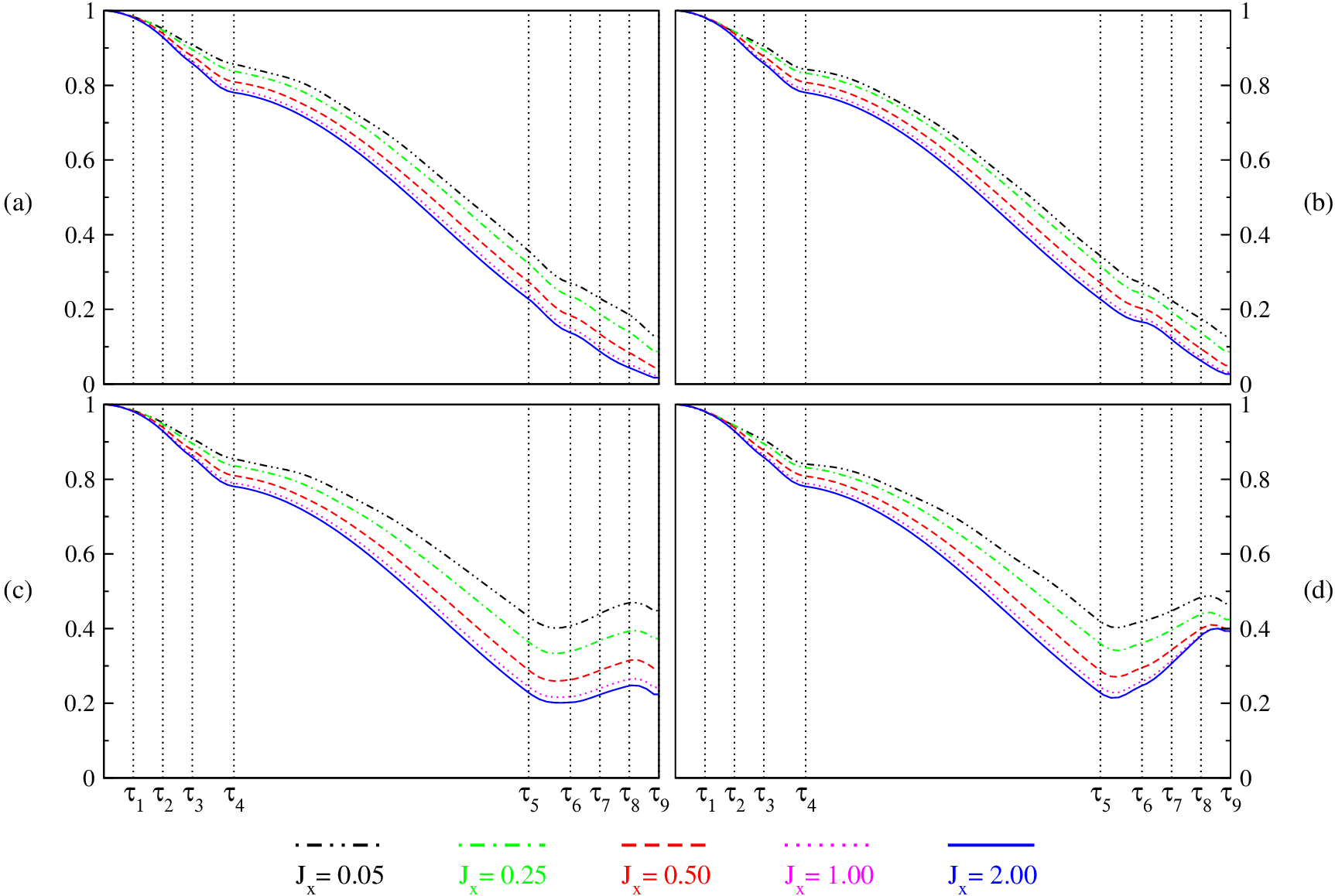}
\caption{(Color online) Fidelity $\mathcal{F}(t)$ vs. time for $xx$ bit--flip errors calculated for initial standard basis states. $\tau_9=1.13$ ns.} 
\label{fid-xx-f}
\end{figure}

\begin{figure}
\centering
\includegraphics[scale=0.35]{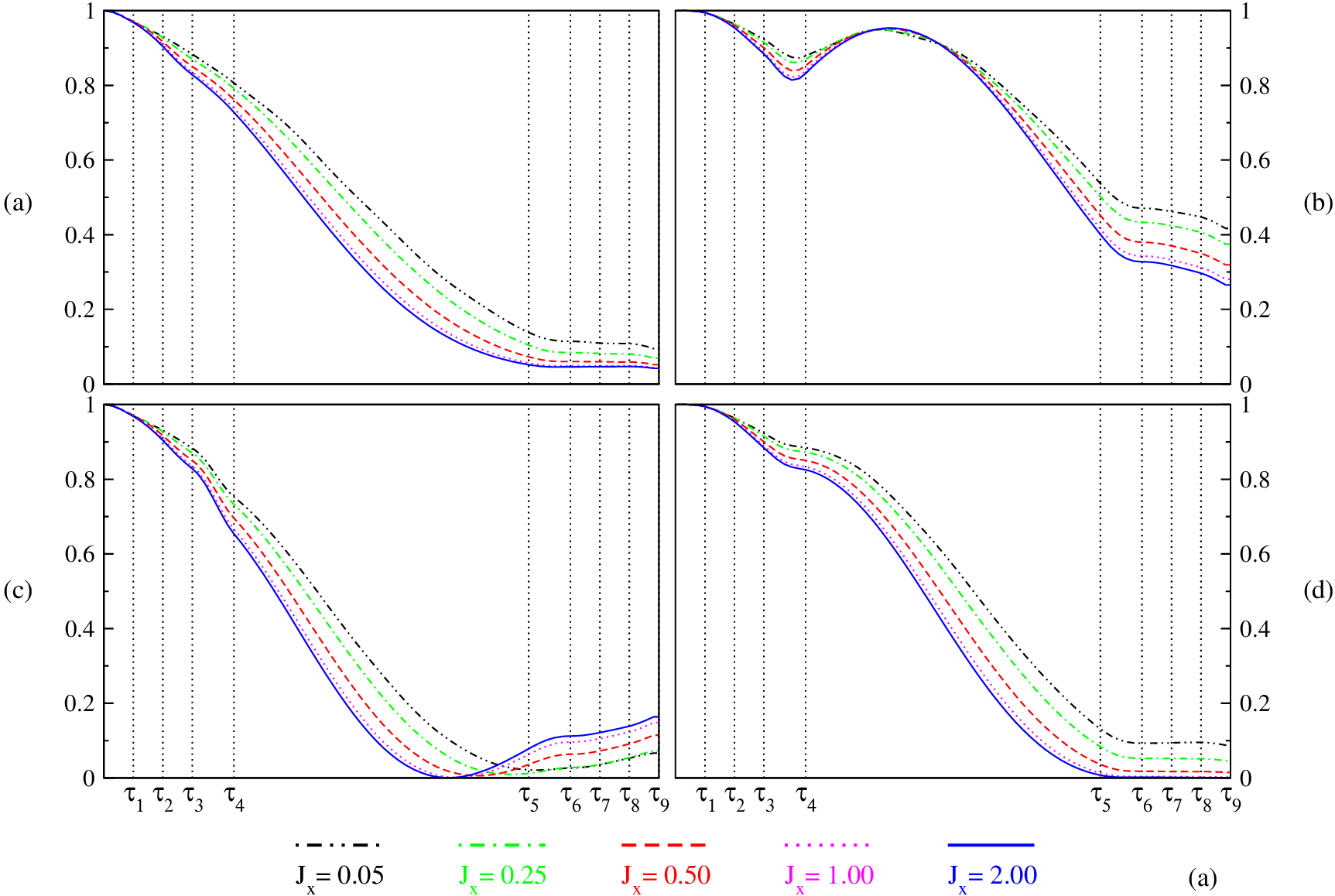}
\caption{(Color online) Fidelity $\mathcal{F}(t)$ vs. time for $xx$ bit--flip errors calculated for initial Bell states. $\tau_9=1.13$ ns.} 
\label{fid-xx-b}
\end{figure}

\begin{figure}
\centering
\includegraphics[scale=0.35]{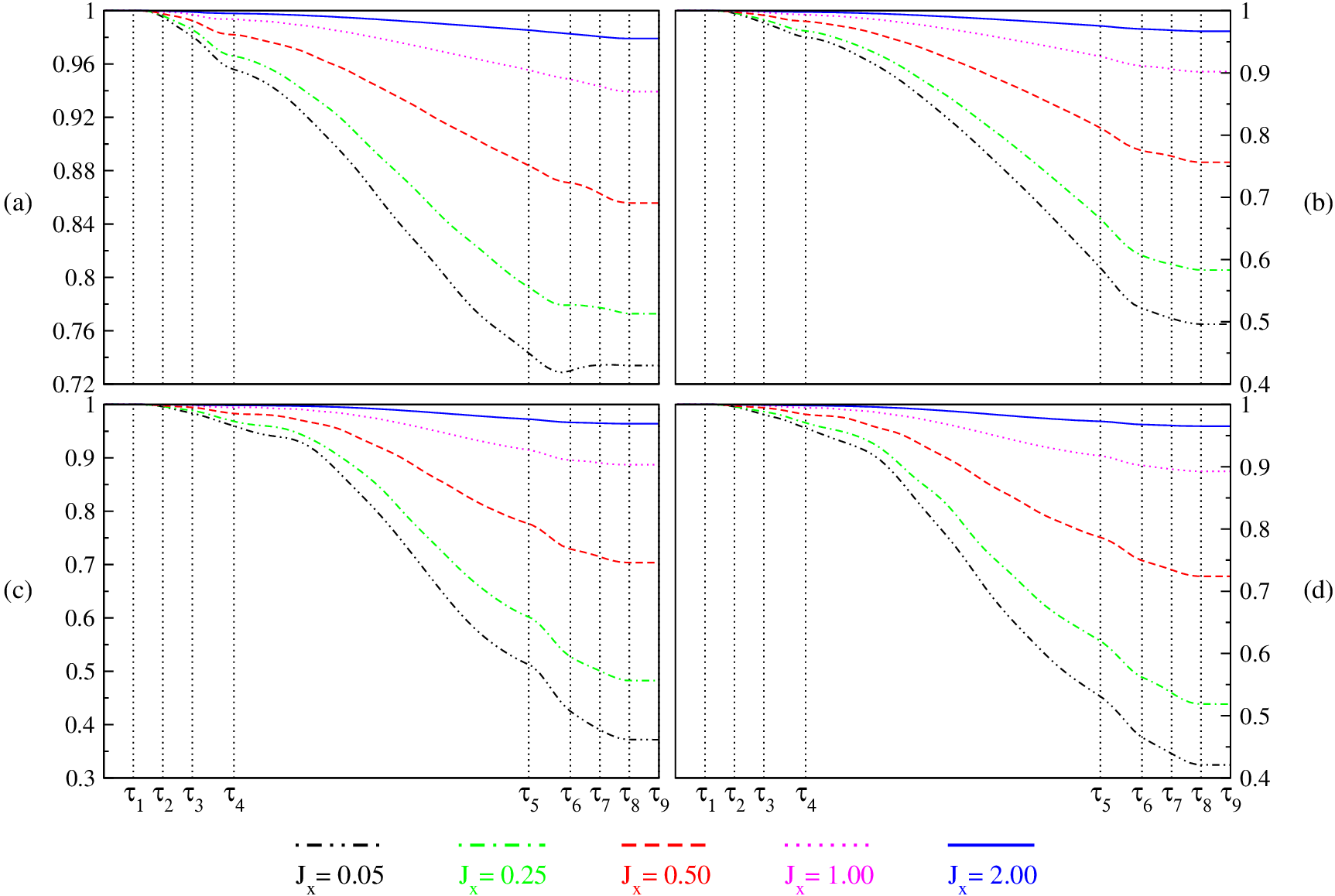}
\caption{(Color online) Fidelity $\mathcal{F}(t)$ vs. time for $zz$ phase errors calculated for initial standard basis states. $\tau_9=1.13$ ns.} 
\label{fid-zz-f}
\end{figure}

\begin{figure}
\centering
\includegraphics[scale=0.35]{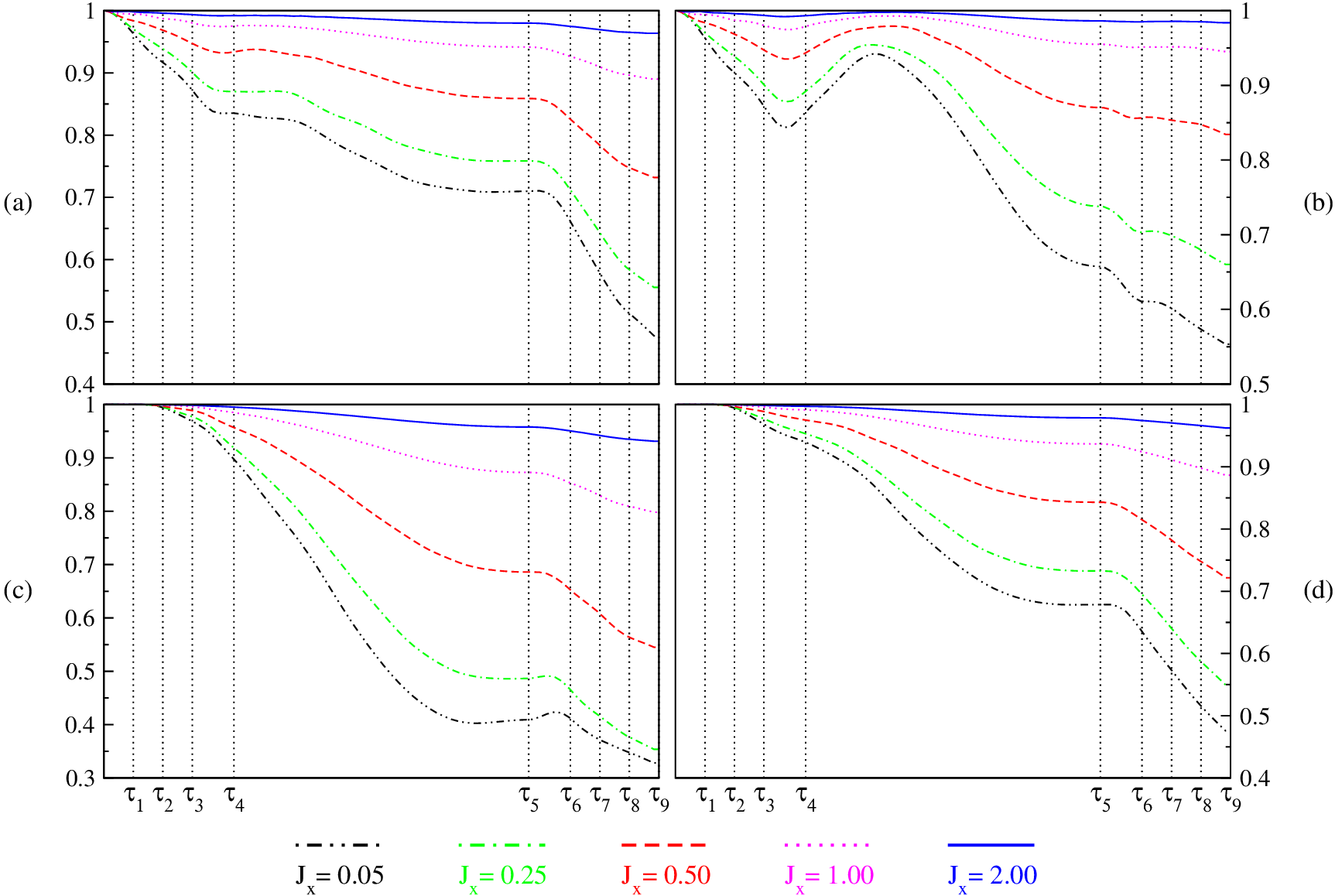}
\caption{(Color online) Fidelity $\mathcal{F}(t)$ vs. time for $zz$ phase errors calculated for initial Bell states. $\tau_9=1.13$ ns.} 
\label{fid-zz-b}
\end{figure}

We plot the fidelity vs time for bit-flip $xx$-type coupling in Figure \ref{fid-xx-f} for standard initial basis states and in Figure \ref{fid-xx-b} for Bell states. For the most part the fidelities are qualitatively similar in that they start out at unity and decay quite uniformly toward zero at the
end of the gate. There is almost no dependence on $J_x$, although fidelity does get slightly worse with increasing $J_x$. Recall, that the 
deviations of the purity from unity were less than ten percent. Here we are losing all fidelity over the course of a single gate. Moreover,
almost all of this error must be of unitary origin since it does not affect the purity. This level of error is disastrous for 
the CNOT gate,  but at least the error is mostly unitary, and thus there may be some systematic way to remove the unitary (error) component by existing error correction schemes \cite{Brown} or specifically tailored new methods.

Some state specific recurrences are seen in \ref{fid-xx-f} (c), (d) and in \ref{fid-xx-b} (c). The behavior of \ref{fid-xx-b} (c) during the two
body gate is also quite odd because there is a complete inversion of the $J_x$ ordered sequence of curves, with the $J_x=0.05$ result going from best to worst
during the two qubit gate.

For phase errors, in Figure \ref{fid-zz-f} for standard initial basis states,  and in Figure \ref{fid-zz-b} for Bell states, we see quite
different behavior. Here the magnitude of the fidelity is quite sensitive to the strength $J_x$ of the intra-bath coupling. Chaotic baths
yield good fidelity while near integrable baths lose sixty percent of the fidelity over the gate. Dependence on initial state is not all
that strong. Note however for Bell states, how (a) and (b), and (c) and (d) show similar behaviors for $t\le \tau_4$.


\section{Discussion of Non-Unitary Effects}

We have seen that increasing $J_x$ results in reduced non-unitary errors. Here we will attempt to explain this effect.
Previous studies have assumed that increasing intra-bath coupling results in increasing chaos\cite{GS,SG,BCS,Tess}. We will show that this
is indeed the case in our model. But our primary concern will be to determine the precise cause of the reduction in errors.
We chose to include $zz$- as well as native $xx$-type coupling in our study even though $zz$-type coupling is not expected in this
particular QC architecture. We will see that suppression of decoherence for these couplings is caused by two different effects.

Subsystem dynamics in the chaotic regime is governed \cite{CW3,CW,SRA} by a master equation of the approximate form
\begin{equation}
\frac{d}{dt}\hat{\rho}_{S}(t)=-(i/\hbar)[\hat{H}_{\rm eff}(t),\hat{\rho}_{S}(t)] -  \int_{0}^{t} dt' W(t-t')\hat{\cal L}_D \hat{\rho}_{S}(t') \label{master}
\end{equation}
where $\hat{H}_{\rm eff}(t)$ is an effective Hamiltonian, including coherent shift terms, the form of which will be discussed in next section, $W(t)$ is a memory function, $\hat{\cal L}_D=(C/\hbar^2)\{ [\cdot\hat{S},\hat{S}]+ [\hat{S},\hat{S}\cdot]\}$ is a dissipative Lindblad-Kossakowski superoperator \cite{dsg}, $\hat{S}$ is the system coupling operator, and $C$ is the canonical variance of the bath coupling operator $\hat{B}$. This equation predicts that decoherence and dissipation in the chaotic
regime are governed by two factors; the variance $C$ of the bath coupling operator, and the positive Gaussian shaped memory function $W(t)$ (unity at $t=0$). Thus, based on this master equation, the suppression of decoherence must either
be governed by a decreased variance or by a more Markovian dynamics.

The more chaotic a bath is, the easier and quicker it can relax, and hence perturbations from a subsystem are quickly thermalized, and subsystem dynamics 
is thus more Markovian. Thus, we expect the memory function to shift its weight to shorter times as $J_x$ increases. As a consequence, the bath will
tend to cause less decoherence. One would also expect the variance in the native bath coupling to decrease with increasing $J_x$ due to the 
vanishing of off-diagonal matrix elements\cite{CW4}. However, this need not be the case with the non-native $zz$ coupling, since it does not
commute with the $xx$ coupling operator. Thus, we will see that both factors favor a reduction of decoherence in the $xx$ case, but reduction
of decoherence in the $zz$ case is entirely due to the increasing Markovity of the dynamics. This subtlety was not fully appreciated in earlier
studies\cite{Tess,CW2}.

To verify these conclusions we begin by showing that the dynamics does indeed become more chaotic with increasing $J_x$. To accomplish that we examine
the bath level spacing statistics and the Lodschmidt echo. Next we examine the variances of the bath coupling operators.
Finally, we look at the product of the variances and memory functions associated with $xx$ and $zz$ couplings.

\subsection{Level statistics and Lodschmidt Echo}

A convenient way to verify the 
crossover from the integrable to chaotic regimes is to observe the nearest neighbor level spacing distribution $P(s)$. As chaos emerges the functional form changes from the Poisson distribution $P(s)=\exp{(-s)}$ characteristic of 
 integrable (i.e. uncoupled) systems to the Wigner--Dyson form $P(s)=(\pi/2)s \exp{(-\pi s^{2}/4)}$ appropriate for chaotic systems.

To verify that this transition does indeed occur in the QC core we performed a level statistics analysis on the unfolded 
spectrum of the 200 lowest eigenenergies of $\hat{H}_{B}$. The unfolded energies $\bar{E}_i$ were generated from actual energies $E_i$
using the smoothed staircase functions $\bar{N}(E)$ via $\bar{E}_i=\bar{N}(E_i)$. Here $\bar{N}(E)$ was obtained as a polynomial
least squares fit to the actual staircase function $N(E)=\sum_{i=1}^{200}\Theta(E-E_i)$ where $\Theta(x)$ is the Heaviside step function.

\begin{figure}[t]
\centering
\includegraphics[scale=0.4]{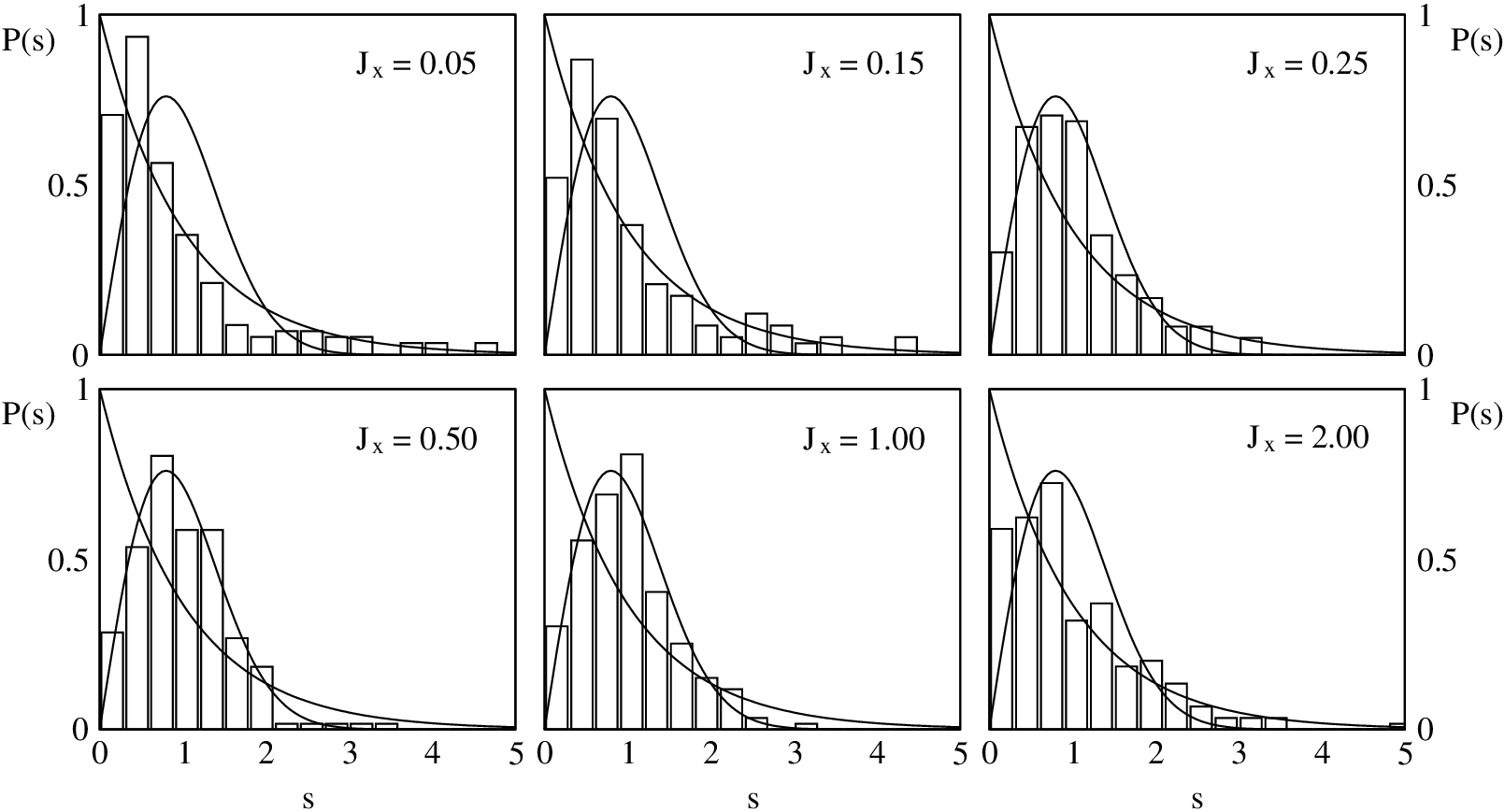}
\caption{Spacing distribution $P(s)$ vs. $s$.}
\label{spacing} 
\end{figure}

A summary of the results of the nearest neighbor spacing analysis are given in Fig. (\ref{spacing}). The onset of chaos can 
be easily seen for a relatively weak inter--qubit coupling strength of $J_{x}=0.15$. Above this value chaos sets in and 
the eigenstatistics are basically consistent with the level-repulsion characteristic of 
Wigner--Dyson statistics. 

While level spacing statistics are considered to be a universal indicator of quantum chaos, they do not provide information on the degree of chaoticity. 
Therefore, we also examined the  Loschmidt echo $M(t)$\cite{Peres}, which is widely believed to be an efficient indicator of quantum chaos\cite{Emerson}, and
which also gives a quantitative indication of the strength of the chaos (i.e., it is similar to a Lyapunov exponent). 

\begin{figure}[t]
\centering
\includegraphics[scale=0.4]{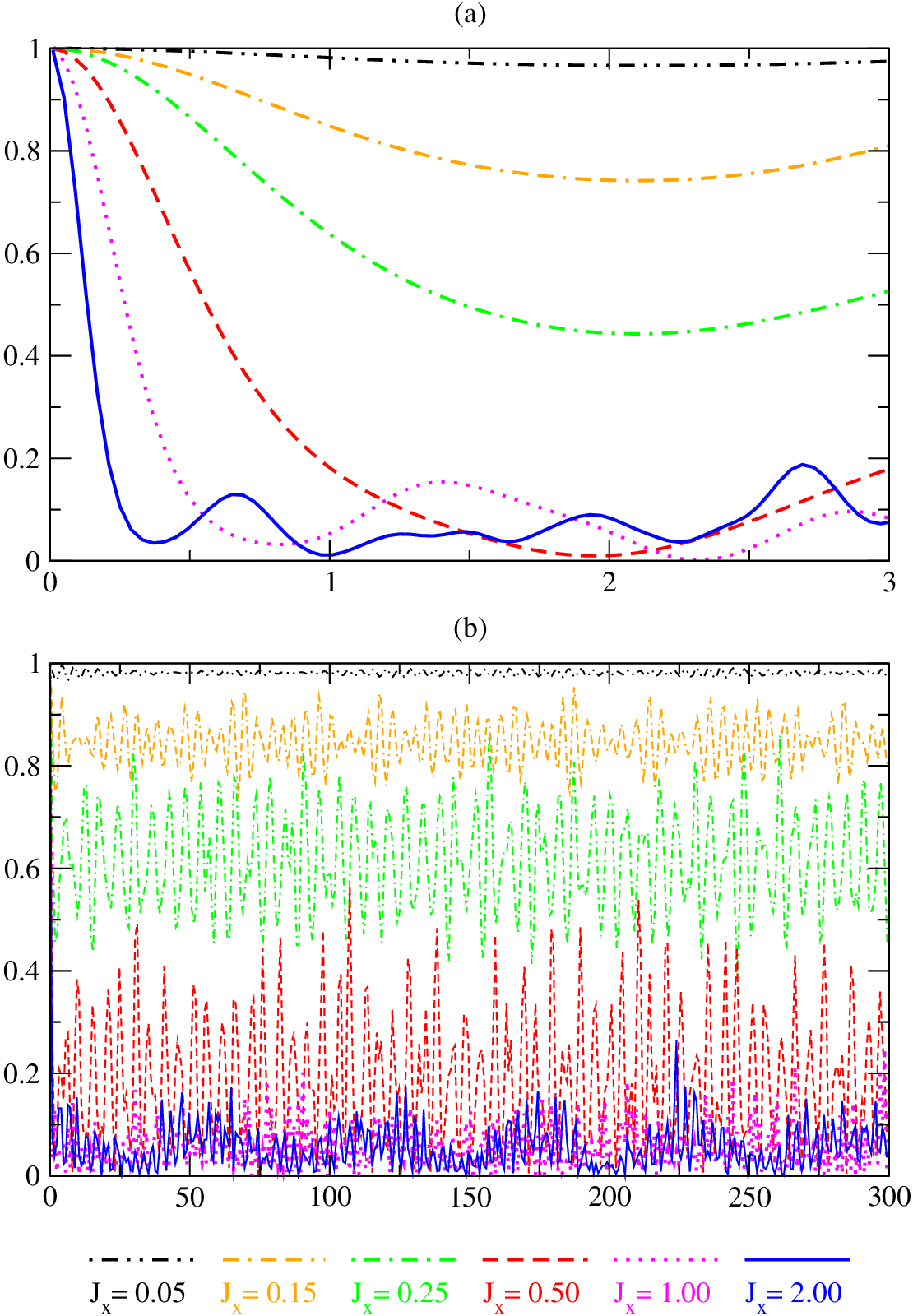}
\caption{(a) Short and (b) long time Loschmidt echo dynamics of bath Hamiltonian for $J_{x}=0.05, 0.15, 0.25, 0.50, 1.00, 2.00$. Time is in units of $\hbar/\epsilon=.038$ ns.}
\label{echo}
\end{figure}

We calculated $M(t)$ for the bath Hamiltonian with the following formula
\begin{equation}     
M(t)=| \langle \psi_{0}| \exp{\{ i\hat{H}_{0}t/\hbar \}}\exp{\{-i(\hat{H}_{0}+\hat{V})t/\hbar\}}|\psi_{0}\rangle|^{2}
\end{equation}
where $|\psi_{0}\rangle$ is the ground eigenstate of $\hat{H}_0$, $\hat{H}_{0}$ is the integrable bath Hamiltonian (i.e. $\hat{H}_B$ for $J_{x}=0.00$) and $\hat{V}$ is the chaos generating perturbation Hamiltonian (i.e., the $xx$ coupling terms) for $J_{x}=0.05,0.15, 0.25, 0.50,1.00,2.00$. A summary of our $M(t)$ calculations is presented in Fig. (\ref{echo}).  
It is clear from Fig. (\ref{echo}) that an increasing magnitude of intra-bath coupling $J_{x}$ results in faster exponential decay of $M(t)$, and systematic approach toward a decay rate insensitive to $J_x$, and this is customarily interpreted as an increasing degree of chaoticity\cite{Peres, Emerson, Gorin}. Note that for smaller $J_{x}$ (i.e $J_x < 0.50$) the echo $M(t)$ does not reach zero.   

\subsection{Canonical variances of bath coupling operators}

\begin{figure}[t]
\centering
\includegraphics[scale=0.3]{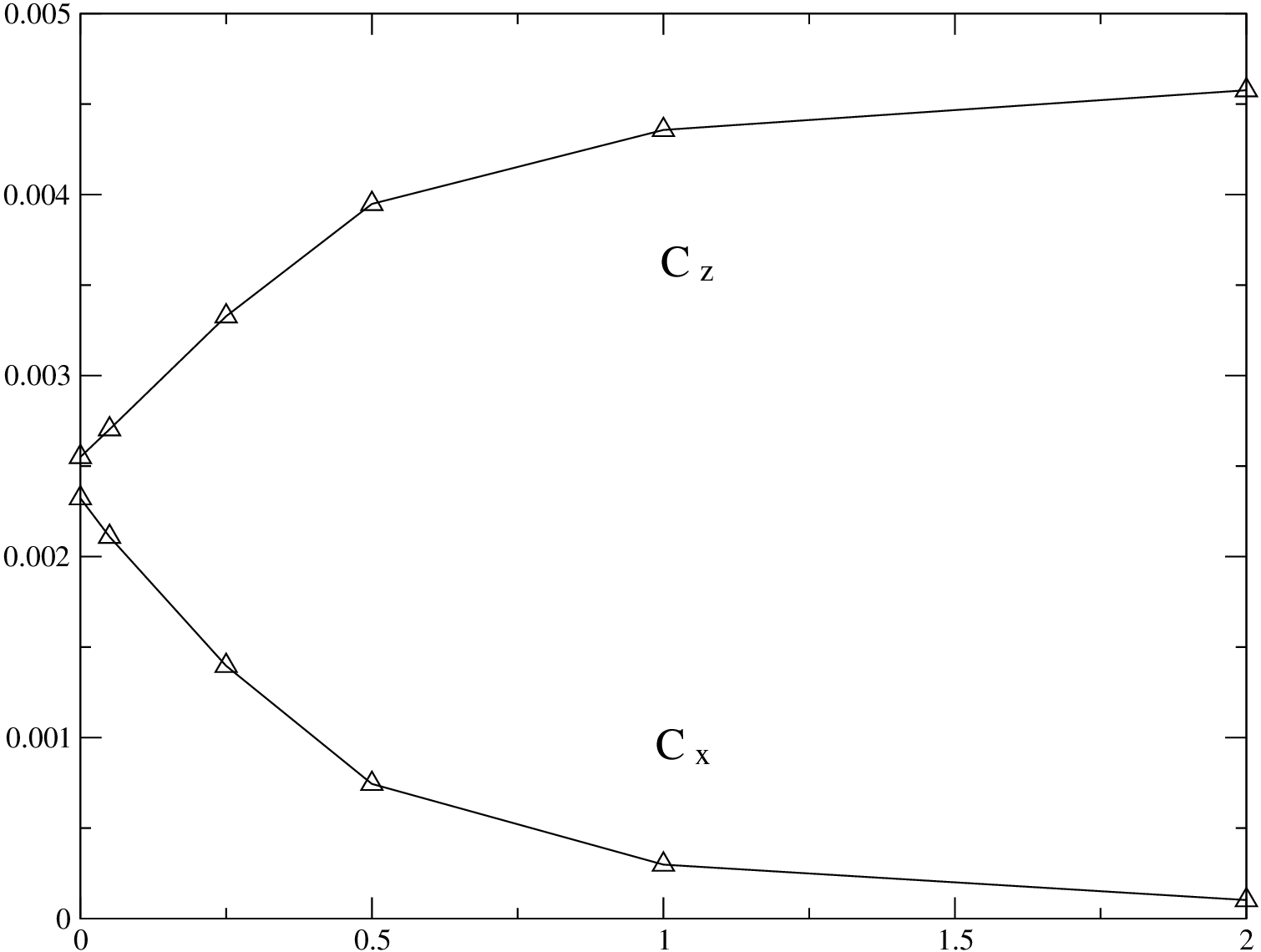}
\caption{Canonical variances in the bath coupling operators for $xx$ and $zz$ couplings vs. $J_x$.}
\label{var}
\end{figure}

For $xx$ coupling we define $\hat{\Sigma}_{x}=\sum_{i=3}^{N+2}\lambda_{i}^{x}\hat{\sigma}_x^i$ and for $zz$ coupling we define
$\hat{\Sigma}_{z}=\sum_{i=3}^{N+2}\lambda_{i}^{z}\hat{\sigma}_z^i$. The variances of these coupling operators can now be defined via
$C_{x}={\rm Tr}_{B}\{ ( \hat{\Sigma}_{x}-\bar{\Sigma}_{x})^{2} \hat{\rho}_{B}(0)\}$ for the $xx$ case, and 
$C_{z}={\rm Tr}_{B}\{ ( \hat{\Sigma}_{z}-\bar{\Sigma}_{z})^{2} \hat{\rho}_{B}(0)\}$ for the $zz$ case. Here 
$\bar{\Sigma}_{x}={\rm Tr}_{B}\{ \hat{\Sigma}_{x} \hat{\rho}_{B}(0)\}$ and $\bar{\Sigma}_{z}={\rm Tr}_{B}\{ \hat{\Sigma}_{z} \hat{\rho}_{B}(0)\}$ denote the canonical
averages. In Figure \ref{var} we plot canonical variances of both bath coupling operators.

Figure \ref{var} shows a decline of the variance for $\hat{\Sigma}_{x}$ with increasing $J_x$. Note however, that there is a 
growth of variance for $\hat{\Sigma}_{z}$ with increasing $J_x$. The decline of $C_x$ with increasing $J_x$ is expected because 
the chaos generating interactions, parameterized by $J_x$, and the bath coupling operator $\hat{\Sigma}_x$ are of the same kind. For strong $J_x$, the eigenstates of $\hat{H}_B$ are also eigenstates of $\hat{\Sigma}_x$. Hence, the off-diagonals of $\hat{\Sigma}_x$ in basis of $\hat{H}_B$ are vanishing. Note that this situation does not require a large thermodynamic bath dimension because of the orthogonality of eigenstates. In parallel to the $\hat{\Sigma}_{x}$ case, a growth of $C_{z}$ with $J_{x}$ can also be understood because the variances are calculated over the same bath states and $\hat{\Sigma}_{x}$ and $\hat{\Sigma}_{z}$ operators are related by canonical commutation rules. 

\subsection{Variance times memory function}

\begin{figure}[t]
\centering
\includegraphics[scale=0.4]{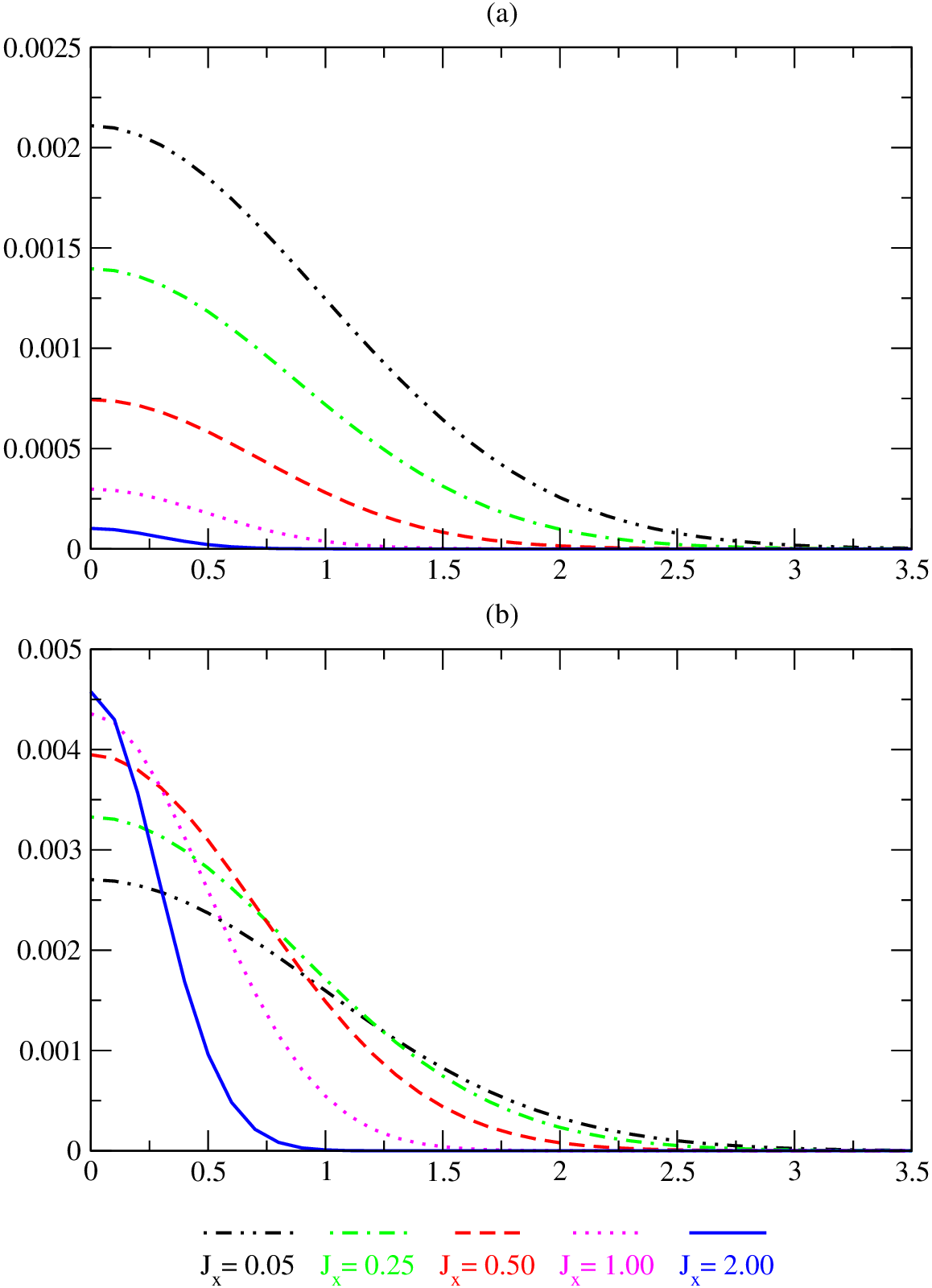}
\caption{Canonical variances times memory function vs. time: (a) $\hat{\Sigma}_{x}$ coupling and (b) $\hat{\Sigma}_{z}$ coupling. Time is in units of $\hbar/\epsilon=.038$ ns.}
\label{varmemo}
\end{figure}

In Figure \ref{varmemo}(a) we plot the variance times the memory function for $\hat{\Sigma}_{x}$ coupling. The dominant effect here is the decrease in the magnitude caused
by the reduction of variance. The function is also however becoming more Markovian, i.e. it is weighted over a smaller time interval.
In Figure \ref{varmemo}(b) we see a growth in the initial magnitude which corresponds to an increase in the variance. But we also see a marked shift toward shorter
times. Again the dynamics is becoming more Markovian with increasing $J_x$, and it is this which causes the reduction of decoherence.
Thus, both types of coupling show a reduction of decoherence in the chaotic regime, but the manifestation of this effect is a bit different.

\section{Discussion of unitary effects}

All unitary effects in our study arise as a consequence of the coherent shift which is an inherent property of Nakajima-Zwanzig type master equations. The coherent shift enters Eq. (\ref{master})
through the effective system Hamiltonian $\hat{H}_{\rm eff}(t)$ which is of the form 
\begin{equation}
\hat{H}_{\rm eff}(t)=\hat{H}_{S}(t)+\hat{S} \bar{B},\label{shift}
\end{equation}
where $\bar{B}$ is the canonical average of the bath coupling operator $\hat{B}$. The second term in Eq. (\ref{shift}) is the 
shift in question, which suggests that a non-negligible contribution from 
the coherent shift should always be expected whenever the canonical average of the subsystem-bath coupling operator has a non-vanishing value (i.e., $\bar{B}\ne 0$). In some of
the older spin-boson and boson-boson studies the existence of coherent shift was not discussed since the coupling operators are of Jaynes-Cummings or coordinate type for
which $\bar{B}=0$, and the shift therefore vanishes. The coherent shift has important consequences when the native subsystem Hamiltonian does not commute with the shift Hamiltonian i.e. $[\hat{H}_S,\hat{S}] \ne 0$. In this case, the effect of the shift corresponds to a distortion of the subsystem dynamics which can cause large unitary errors. In cases where the subsystem and shift Hamiltonians commute, the coherent shift, more or less, corresponds to an energy shift similar to Lamb shift like contributions. In this case, generation of unitary errors may be more easily avoided.   

\begin{figure}[t]
\centering
\includegraphics[scale=0.3]{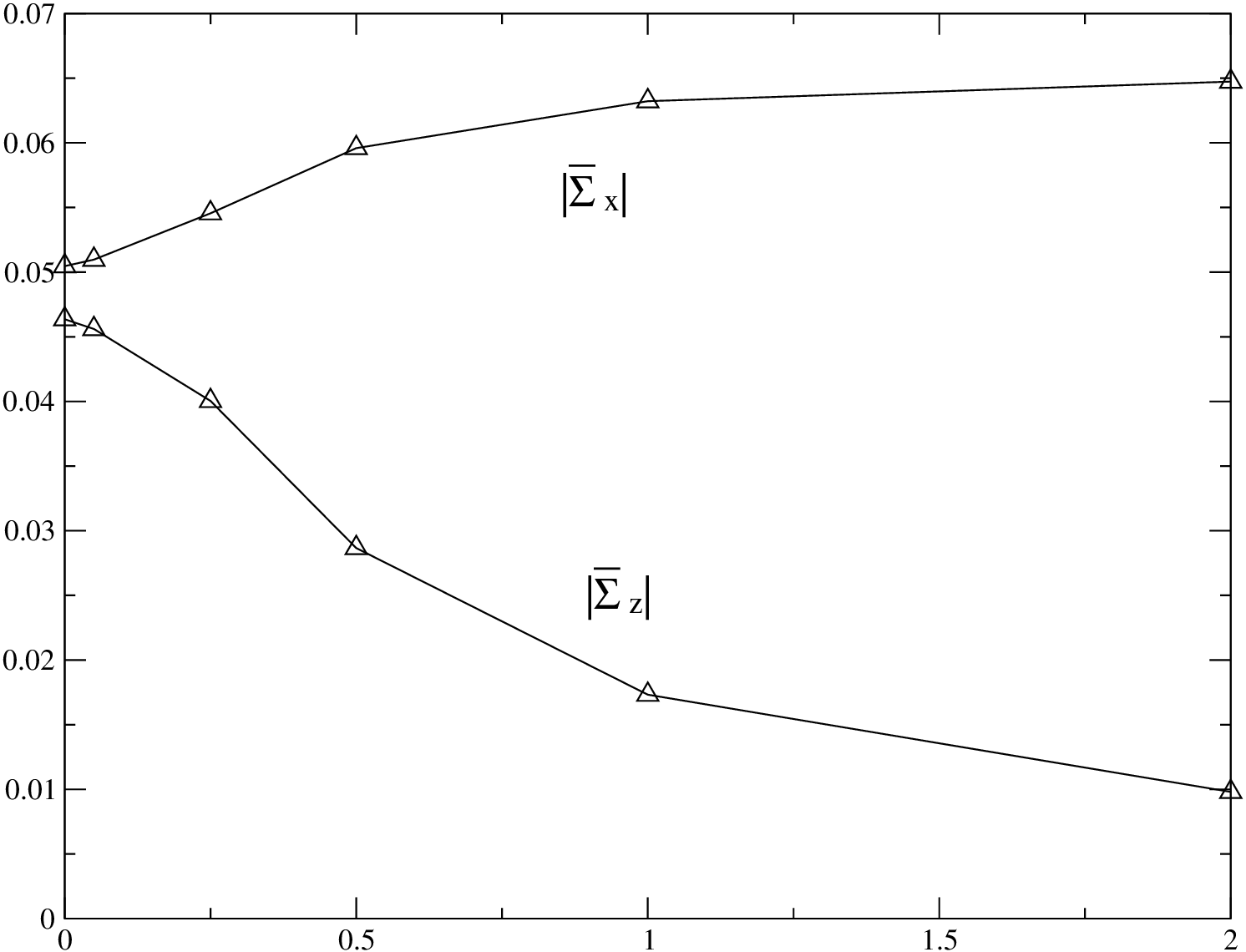}
\caption{$|\bar{\Sigma}_{x}|$ and $|\bar{\Sigma}_{z}|$ plotted vs. $J_{x}$ in units of $\epsilon$.}
\label{bars}
\end{figure}

In Figure (\ref{bars}) we plot the absolute values of the canonical averages of the bath-coupling operators i.e. $|\bar{\Sigma}_{x}|=|{\rm Tr}\{ \hat{\Sigma}_{x} \hat{\rho}_{B}(0)\}|$ and $|\bar{\Sigma}_{z}|=|{\rm Tr}\{ \hat{\Sigma}_{z} \hat{\rho}_{B}(0)\}|$ vs. increasing values of $J_{x}$. In the case of phase errors, an overall decrease in the average is seen with increasing magnitude of $J_{x}$. In the case of bit-flip errors, however, the increasing magnitude of $J_{x}$ results in a small amount of increase in the average. This is clearly consistent with the observed improvement in fidelity with increasing $J_{x}$ for phase type coupling (see Figures \ref{fid-zz-f} and \ref{fid-zz-b}) and the slight decline of fidelity for bit-flip type coupling (see Figures \ref{fid-xx-f} and \ref{fid-xx-b}). 

The unitary effects observed in our calculations are not of Lamb shift type and in fact are quite worrying. The magnitude of the errors in
fidelity for the span of a single CNOT gate is much larger than one would have expected based on the results of our previous study\cite{CW3}, where a single qubit subsystem is configured to detect internal bath dynamics. We call this set-up a single qubit Rabi detector \cite{CW3}. The bath
Hamiltonian, and the $xx$ coupling operator and its strength, employed in \cite{CW3} were identical to those used in our CNOT study so that the magnitude of the shift is not altered, but somehow it is dramatically more harmful. Moreover, this has nothing to do
with the dimensionality of the subsystem. 

\begin{figure}[t]
\centering
\includegraphics[scale=0.35]{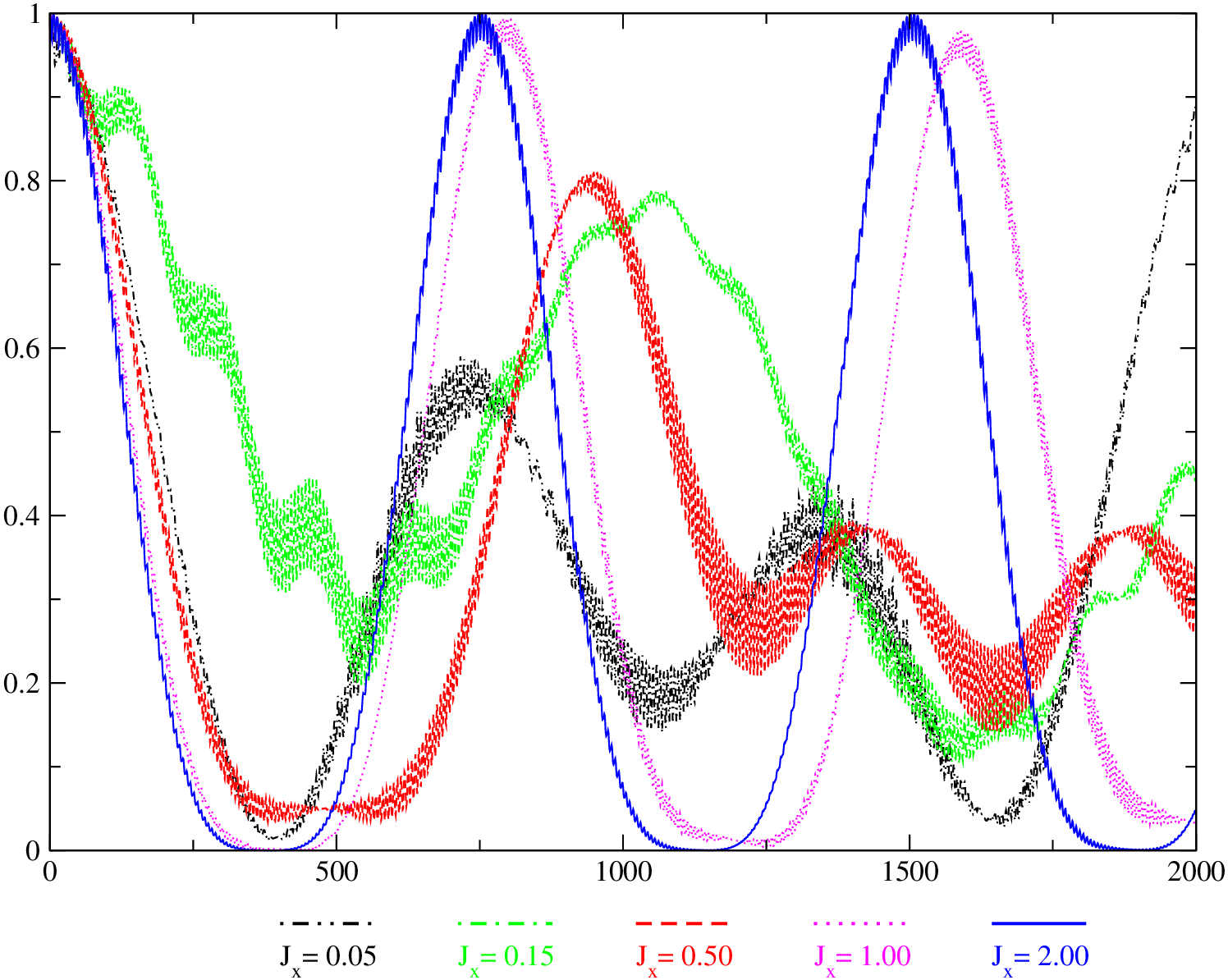}
\caption{Fidelity ${\cal F}(t)$ vs. time for two-qubit Rabi detector. Time is in units of $\hbar/\epsilon=.038$ ns.}
\label{rabi2-fid}
\end{figure}

To show this, we have redone these calculations for a two-qubit subsystem (two-qubit Rabi detector) which has the subsystem Hamiltonian
\begin{equation}
\hat{H}_{S} = -\frac{1}{2} ( {\cal B}_{z}\hat{\sigma}_{z}^{1}+{\cal B}_{z}\hat{\sigma}_{z}^{2}),
\end{equation}
where ${\cal B}_{z}=1$ and the dynamics evolves from an initial state $|\psi_0\rangle=.5(|0\rangle+|1\rangle)\otimes(|0\rangle+|1\rangle)$. We plot the fidelity for the two-qubit Rabi detector in Figure  \ref{rabi2-fid}, which shows that the two-qubit Rabi detector experiences errors in fidelity which are similar to those of the single qubit Rabi oscillator.   

The only remaining possibility for the large errors in fidelity is the state dependency of the CNOT gate. That is, the rapidly changing nature of the state on which the CNOT gate operates is responsible for the large magnitude error. In what follows, a direct analogy between the CNOT subsystem and a kicked top can readily be established by viewing the fidelity (in the absence of the weak non-unitary
effects) as being similar to the Loschmidt echo of a kicked top.

A second unexpected effect is that the fidelity decay seems to be almost independent of $\bar{B}$ which itself changes with $J_x$ as shown
in Figure \ref{bars}. In the single qubit Rabi detector study \cite{CW3} the fidelity decay time was highly sensitive to $J_x$, and the same is true of the two-qubit Rabi
detector (see Fig. \ref{rabi2-fid}). Here the period of the decay increases substantially by 10 ns when $J_x$ increases from zero to .15, then declines from $J_x=.15$ to $J_x=.5$, and finally moves toward some saturated value after $J_x=1$. Fidelity decay times for the CNOT with native $xx$ coupling vary by less than .1 ns. This is thus a major effect. It seems quite likely that these two unusual
effects are somehow related, and that CNOT dynamics has a phenomenology similar to that of the Loschmidt echo of a kicked top\cite{LEKT}.

It is well-known that there are two regimes of Loschmidt echo of a kicked top\cite{LEKT}:  the fast exponential decay regime which is insensitive to the perturbation strength and the Golden Rule regime where decays are slower and decay rates depend on the 
perturbation strength\cite{LEKT}. Our results for the CNOT gate and Rabi detectors also fit into this picture and the origin of the two unexpected unitary effects can be explained as arising from the kicked nature of the CNOT
gate. 

The two-qubit Rabi detector would correspond to vanishingly weak
kicking which would be expected to lie in the Golden Rule regime where decays are slower and decay rates depend on the 
perturbation strength\cite{LEKT}. However, note also that the sensitivity of fidelity to $J_{x}$ vanishes in the strongly chaotic bath regime (i.e. $J_{x}=1$ and $J_{x}=2$) where the period of decay saturates toward a certain value.

On the other hand, it appears that the CNOT gate for $xx$-type coupling case lies in the exponential decay regime where the the fidelity does not show any sensitivity to perturbation strength. However, the CNOT gate for the $zz$-type coupling case lies in the Golden Rule regime where the dominant effect is the high sensitivity of fidelity to the perturbation strength. The kicked-top nature of the CNOT gate suggests that removing the effects of the
decay {\em after } completion of the gate is essentially impossible, which means that error correction strategies for the 
shift must be performed during each subcomponent of the gate.

\section{Summary}
We studied the effects of static one-- and two--body flaws on a CNOT gate performed on
part of a larger Josephson charge-qubit QC \cite{Nori}. We observed reduction of decoherence
with increasing intra-bath coupling, and a unitary shift of the gate, both as a result of residual
interactions with the idle part of the QC. For this architecture it seems that the coherent shifting is far more harmful than 
internal or external decoherence effects. Moreover, the large magnitude of the error in fidelity and its insensitivity to
the strength of the subsystem-bath interaction strength put this QC in the so-called exponential decay regime
of a kicked top. This means that retroactive error correction is impossible, and error correction strategies
will have to operate concurrently with all gate components. We have given clear and complete explanations for all of these effects.

Our results are specific to this particular architecture and parameter regime
and it would be interesting to see similar studies of other architectures.
We anticipate that the observed coherent shift is a universal feature of qubit based flawed QCs. Since the magnitude of the shift we observed is so large, an explicit experimental verification of this interesting effect should be readily achievable for a relatively small number of qubits. 

\ack
The authors acknowledge the support of the Natural Sciences and Engineering Research Council of Canada. This research employed WestGrid computing resources.

\end{document}